# Querying structural and functional niches on spatial transcriptomics data

Mo Chen[1,8,9]†, Minsheng Hao[1]†, Xinquan Liu[3,4]†, Lin Deng[3,4], Peng Liu[1], Chen Li[1], Dongfang Wang[5], Kui Hua[6], Liang Guo[7], Xuegong Zhang[1,2]*, Lei Wei[1]*

[1] MOE Key Laboratory of Bioinformatics and Bioinformatics Division of BNRIST, Department of Automation, Tsinghua University, Beijing, China.

[2] Center for Synthetic and Systems Biology, School of Life Sciences and School of Medicine, Tsinghua University, Beijing, China.

[3] Department of Thoracic Surgery, National Cancer Center/National Clinical Research Center for Cancer/Cancer Hospital, Chinese Academy of Medical Sciences and Peking Union Medical College, Beijing, China.

[4] Peking Union Medical College, Chinese Academy of Medical Sciences, Beijing, China.

[5] Biomedical Pioneering Innovation Center (BIOPIC), Peking University, Beijing, China.

[6] Cancer Research UK Cambridge Institute, University of Cambridge, Cambridge, UK.

[7] Department of Immunology, School of Basic Medical Sciences, Harbin Medical University, Harbin, China.

[8] Zhongguancun Academy, Beijing, China.

[9] Zhongguancun Institute of Artificial Intelligence, Beijing, China.

* Corresponding author: zhangxg@tsinghua.edu.cn (X.Z.), weilei92@tsinghua.edu.cn (L.W.)

† These authors contributed equally to this work.

## Abstract

Cells in multicellular organisms coordinate to form structural and functional niches. With spatial transcriptomics (ST) enabling gene expression profiling in spatial contexts, it has been revealed that spatial niches serve as cohesive and recurrent units in physiological and pathological processes. These observations suggest universal tissue organization principles encoded by conserved niche patterns, and call for a query-based niche analytical paradigm beyond current computational tools. In this work, we defined the niche-query task, which is to identify similar niches across ST samples given a niche of interest (NOI). We further developed QueST, a specialized method for solving this task. QueST models each niche as a subgraph, uses contrastive learning to learn discriminative niche embeddings, and incorporates adversarial training to mitigate batch effects. In simulations and benchmark datasets, QueST outperformed existing methods repurposed for niche querying, accurately capturing niche structures in heterogeneous environments and demonstrating strong generalizability across diverse sequencing platforms. Applied to tertiary lymphoid structures in renal and lung cancers, QueST revealed functionally distinct niches associated with patient prognosis and uncovered conserved and divergent spatial architectures across cancer types. Applied to a combinatorial spatial perturbation dataset, QueST demonstrated a complete *de novo* discovery-oriented workflow, characterizing previously unresolved tumor nodules through querying. These results demonstrate that QueST enables systematic, quantitative profiling of spatial niches across samples, providing a powerful tool to dissect spatial tissue architecture in health and disease.

## Introduction

Cells in multicellular organisms interact and coordinate within tissues, forming spatial niches as localized structural and functional communities[1,2]. These niches, characterized by the composition and spatial arrangement of cells, orchestrate complex biological processes such as immune responses, tissue regeneration, and disease progression[3,4]. Recent advances in spatial transcriptomics (ST) technologies[5–9] have enabled transcriptome-wide, spatially resolved views of tissue architecture, laying the groundwork for data-driven investigations of spatial niches. Recent studies have identified niches with distinct structural and functional characteristics in specific biological contexts[10–13]. Meanwhile, large-scale cross-tissue analyses[14,15] have revealed that tissues are composed of recurrent and functionally specialized multicellular units. These findings suggest that some spatial niches are not unique to individual datasets but may represent conserved building blocks of tissue architecture that transcend individual tissue samples or conditions.

These observations converge on a critical insight: spatial niches are not merely dataset-specific instances but may represent universal principles of tissue organization. This realization brings forth a fundamental question: given a niche of interest (NOI), can we systematically identify similar niches across tissues, individuals, or disease states? We therefore define the "niche-query task": to identify other niches or local areas that share similar spatial characteristics with a given niche of interest across ST datasets. Solving this problem is crucial for uncovering conserved tissue modules, characterizing disease-relevant microenvironments, and generalizing niche-level insights across biological systems. This task is distinct from existing cross-sample analytical paradigms in spatial transcriptomics. Most current approaches are cell-centric, anchoring the analysis on individual cells to describe each cell's local microenvironment or to delineate broad tissue regions by clustering these cell-level descriptions[16,17]. The niche-query task instead treats the niche itself as the unit of analysis, defining it as a small, conserved local structure with characteristic cellular composition and spatial topology that can recur across diverse tissues and disease conditions, even when embedded in heterogeneous tissue backgrounds.

Solving the niche-query task demands systematically comparing structured multicellular local areas across samples. The cornerstone of this task is to establish a robust and informative representation for each niche. Such niche representation should have three key properties: (1) "Central-cell Invariance": All cells within a niche contribute equally, without privileging a central cell; (2) "Strict Locality": The representation is determined solely by cells within the niche boundary; (3) "Structure Awareness": The representation captures information not only of cell composition but also of spatial organization, enabling discrimination between niches with similar composition but different topology. From this perspective, existing representation learning methods in ST[18–26] largely follow a cell-centric paradigm, where embeddings are learned at the level of individual cells and spatial context is incorporated via message passing. While such approaches work in clustering-oriented scenarios where some methods[16,17,27–29] use context-enriched cell embeddings to approximate niche-level features, the underlying representation remains centered on individual cells. As a result, these methods do not explicitly enforce locality at the niche boundary and do not directly encode internal spatial structure at the level of the niche as a whole, and thus are not suited for querying niches across diverse tissue backgrounds. Other studies employ pooling strategies over node embeddings[30–32] to generate niche embeddings, but hardly capture the structural intricacies of niche organization. Besides, biologically similar niches may exist in highly heterogeneous spatial environments[33], and batch effects further complicate cross-sample niche comparison[34]. Without resolving these issues, spatial niche analysis lacks a systematic foundation and cannot yield broadly applicable biological insights.

To address these challenges, we developed QueST, a method for niche representation learning and querying across multiple samples. QueST models a niche as a subgraph of cells and employs a subgraph contrastive learning strategy to satisfy the three key principles. In the strategy, positive and negative niche pairs are constructed by partially shuffling the spatial proximity graph while keeping certain parts intact. To mitigate batch effects across samples, QueST incorporates an adversarial training scheme inspired by Generative Adversarial Networks[35], encouraging the model to learn sample-invariant niche representations.

We first demonstrated that QueST can effectively capture the structural characteristics of spatial niches in highly heterogeneous environments using a simulated ST dataset. We then systematically benchmarked QueST on real ST datasets against existing ST-analysis methods repurposed for the niche-query task, first on a multi-sample cortical dataset and further across different sequencing technologies and spatial resolutions. Beyond benchmarking, we applied QueST to analyze tertiary lymphoid structures (TLS) within highly heterogeneous tumor microenvironments (TME). We identified both shared and distinct characteristics among TLS across different samples, tissues, and even cancer types profiled with different sequencing platforms. Finally, we applied QueST to a combinatorial spatial perturbation dataset and demonstrated a complete discovery-oriented workflow, identifying NOI *de novo* and characterizing previously unresolved tumor nodules through querying. Together, these results highlight QueST as a powerful and versatile tool for spatial niche representation and comparison, with its niche representation serving as the key entity for encoding the structural and functional information of niches.

# Results

## Study overview

We define the niche-query task in spatial transcriptomics (ST) data as the problem of identifying spatial niches that exhibit similar spatial architecture and gene expression characteristics to a specified niche of interest (NOI). In each ST sample, a spatial niche is defined as a subgraph centered on an individual cell (Fig. 1a). A dataset comprising multiple tissue samples therefore contains a diverse collection of such spatial niches. Given an NOI from one sample (referred to as the source sample), the task is to query across the remaining samples (referred to as target samples) to find similar candidate niches (Fig. 1b). We elaborated on the distinctions between this task and previous analytical paradigms in spatial transcriptomics in Supplementary Note 1.

To solve this problem, we developed QueST, a computational method that learns niche representations through a unified training process. QueST models ST data as spatial graphs, with nodes representing individual cells with normalized gene expression, and edges connecting spatially adjacent cells. Each niche is defined as a *K*-hop subgraph centered at a cell. The goal of QueST is to encode the niche's expression and structural information into an embedding via the combination of reconstruction-based self-supervised learning and contrastive learning.

QueST's model architecture comprises three components: a graph autoencoder (GAE), a contrastive learning (CL) module, and an adversarial batch removal module. The GAE (Fig. 1c) learns latent node embeddings by reconstructing gene expression. These embeddings are then aggregated by mean pooling within each *K*-hop subgraph to generate the corresponding niche embedding. To enhance the ability to encode structural and compositional variations, the CL module (Fig. 1d) generates positive and negative niche pairs by partially shuffling the graph (Methods). Positive pairs consist of the same niche from the original and corrupted graphs, enabling the model to identify similar niches under heterogeneous backgrounds. Negative pairs differ in cell composition and spatial topology, encouraging the model to distinguish dissimilar niches. To further improve generalizability, QueST employs adversarial training (Fig. 1e) to reduce batch effects: a batch discriminator is trained to distinguish batches from the niche embeddings, while the encoder in GAE is trained to deceive this batch discriminator. We discussed the design rationale of these modules in Supplementary Note 2.

During inference, QueST scans through the target ST samples for *K*-hop subgraphs as candidate niches and computes embeddings for them. Given a query NOI, QueST measures its similarity to candidate niches using the cosine similarity of their niche embeddings, yielding a predicted niche matching score, and retrieves similar niches based on this score (Fig. 1f).

We compared QueST with state-of-the-art representation learning methods, including GraphST[21], STAGATE[18], NicheCompass[16], CAST[22] and SLAT[24]. Among them, GraphST and STAGATE were designed for spatial domain detection. NicheCompass was for niche clustering and identification. CAST and SLAT were developed for spatial alignment, which is a task closely related to niche querying, as both tasks aim to identify similarity patterns across samples. Since these methods are not designed for the niche-query task, we used their node embeddings as proxy niche embeddings centered at each corresponding node and computed cosine similarity between niche embeddings as their predicted matching score. For all methods, we computed the cosine similarity between the embedding of the query NOI and those of all candidate niches to derive the predicted niche matching scores.

As an additional reference, we included a Region Matching (RM) approach that computes niche similarity based on subgraph topology and annotation labels of cells/spots (Methods). We evaluated RM in two settings: using ground truth annotations (RM-Ideal) and using unsupervised cluster labels derived from GraphST (RM-GraphST). Among them, RM-Ideal was used as the gold standard to benchmark the performance of different methods throughout most subsequent analyses, as it approximates true niche similarity when annotations are in proper granularity and perfectly accurate. Though, it is important to note that the RM-Ideal approach is not practical for real applications, as it lacks computational efficiency due to being difficult to parallelize, and is highly sensitive to the accuracy of input annotation labels, as we demonstrate later in this study. Therefore, although RM-Ideal provides a valuable theoretical upper bound, it should not be considered a viable solution to the niche-query task.

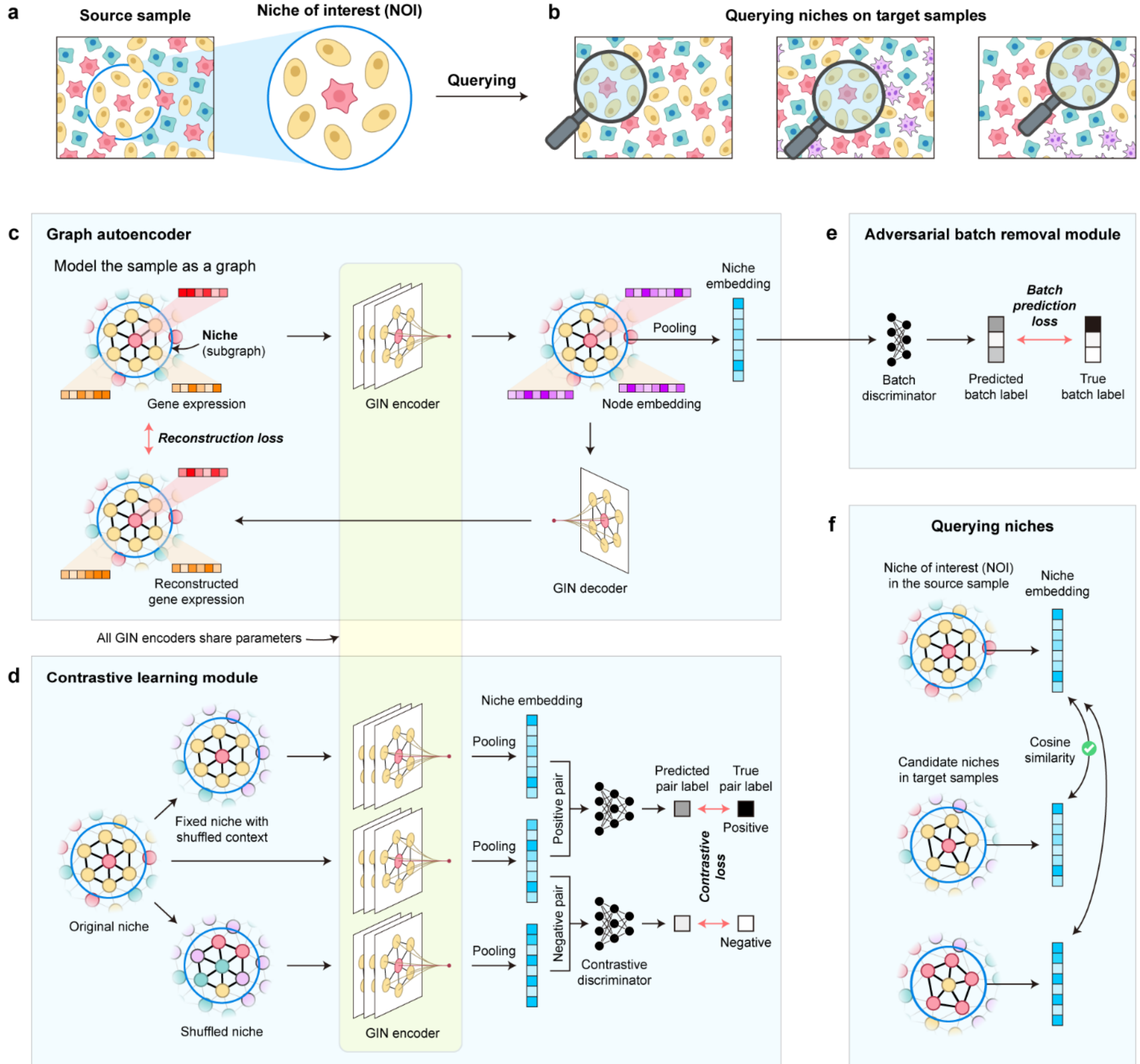

**Fig. 1. Overview of the niche-query task and QueST algorithm. a.** Illustration of the definition of a spatial niche. **b.** Schematic explanation of the niche-query task. **c**. Graph autoencoder (GAE) module of QueST. Given the ST data, QueST models cells/spots as nodes, and niches as *K*-hop subgraphs. The GAE module takes normalized gene expression and the spatial graph as input, reconstructs expression data as the output, and learns latent node embeddings in this process. The node embeddings are pooled within each niche as the initial niche embeddings. **d**. Contrastive learning module of QueST. This module refines niche embeddings by discriminating positive and negative niche pairs and forces the model to encode structural and compositional information in niche embeddings. **e**. Adversarial learning module of QueST. The encoder is trained against an MLP batch discriminator and is encouraged to output

batch-free niche embeddings in the process. **f**. Illustration of the niche querying process. QueST computes cosine similarity between niche embeddings as the predicted niche matching score, and uses it for niche querying.

## Querying niches across topological and environmental variations

To assess whether QueST captures niche-level structural differences under heterogeneous conditions, and to concretely illustrate the niche-query task itself, we designed a deliberately minimal simulation that provides a transparent and controlled setting for evaluation. Specifically, we simulated an ST dataset with two samples using scCube[36] (Fig. 2a). Sample 1 (S1) was composed primarily of endothelial cells, serving as the background tissue in which several distinct niches were embedded. In contrast, Sample 2 (S2) used macrophage/monocyte cells as the background tissue. Within each sample, we embedded three types of NOIs, which varied in both cellular composition and spatial topology: the T-core niche, with T cells at the center and cancer/epithelial cells at the periphery; the T-edge niche, with the reverse structure; and the B-edge niche, which replaced peripheral T cells with B cells.

We introduced three niche-query tasks, T-core querying, T-edge querying, and B-edge querying, each aiming to identify all matched candidate niches in the target sample S2, given a query NOI from the source sample S1 (Fig. 2a). Fig. 2b-d and Fig. S1 showed the spatial distribution of predicted matching scores for each method. RM-Ideal achieved the best performance, owing to its access to ground-truth annotations, whereas RM-GraphST performed substantially worse due to the use of imperfect labels. QueST closely matched the performance of RM-Ideal, accurately distinguishing between T-core and T-edge niches and assigning consistently low similarity to unrelated B-edge niches in S2. In contrast, the other methods all struggled to distinguish between structurally similar but topologically distinct niches. For example, alignment methods such as PASTE and STAligner, as well as foundation models such as Nicheformer and Novae, yielded almost identical results for T-core and T-edge queries. Most of the existing methods could not separate these T-core and T-edge niches from B-edge niches either, highlighting their limitations in handling niche differences in highly heterogeneous spatial contexts.

We first evaluated niche querying performance by framing it as a binary classification task,

where niches of the same type as the query were treated as positive cases. Model performance was evaluated using the area under the precision-recall curve (AUPRC). As shown in Fig. 2e, QueST achieved the highest accuracy in identifying the correct target niches for all querying types, with AUPRC values approaching 1.0, consistent with the spatial patterns observed in Fig. 2b-d.

Since the variation in niche composition and topological structure is continuous, we further evaluated each method's ability to capture graded similarity by computing the Pearson correlation coefficient (PCC) between predicted matching scores and RM-Ideal (Fig. 2e). QueST again showed superior performance, with PCC values above 0.8 for all query types. In contrast, RM-GraphST exhibited markedly lower correlation, underscoring the sensitivity of RM-based approaches to the accuracy of input annotations.

In previous studies, spatial alignment methods[22–26] were developed to identify similar spatial patterns across samples. However, spatial alignment and niche querying differ fundamentally in both modeling unit and objective. Spatial alignment methods operate primarily at the level of individual cells and are optimized for correspondence rather than selective retrieval. As a result, they do not explicitly model niches as integrated spatial units, and therefore struggle to distinguish regions that share similar cellular composition but differ in higher-order spatial organization. We illustrate this with the T-edge query (Supplementary Note 3, Fig. S2), where the two alignment methods, CAST and SLAT, encountered extensive misalignment due to the fundamental limitation of spatial alignment, but QueST correctly retrieved T-edge niches. These results highlight a key distinction: spatial alignment focuses on establishing correspondences of cells across samples, whereas niche querying requires identifying spatially organized cell populations as cohesive units.

In summary, our simulation results demonstrate that QueST robustly captures niche-level differences under substantial topological and environmental heterogeneity, outperforming existing methods in both classification-based and similarity-based evaluations.

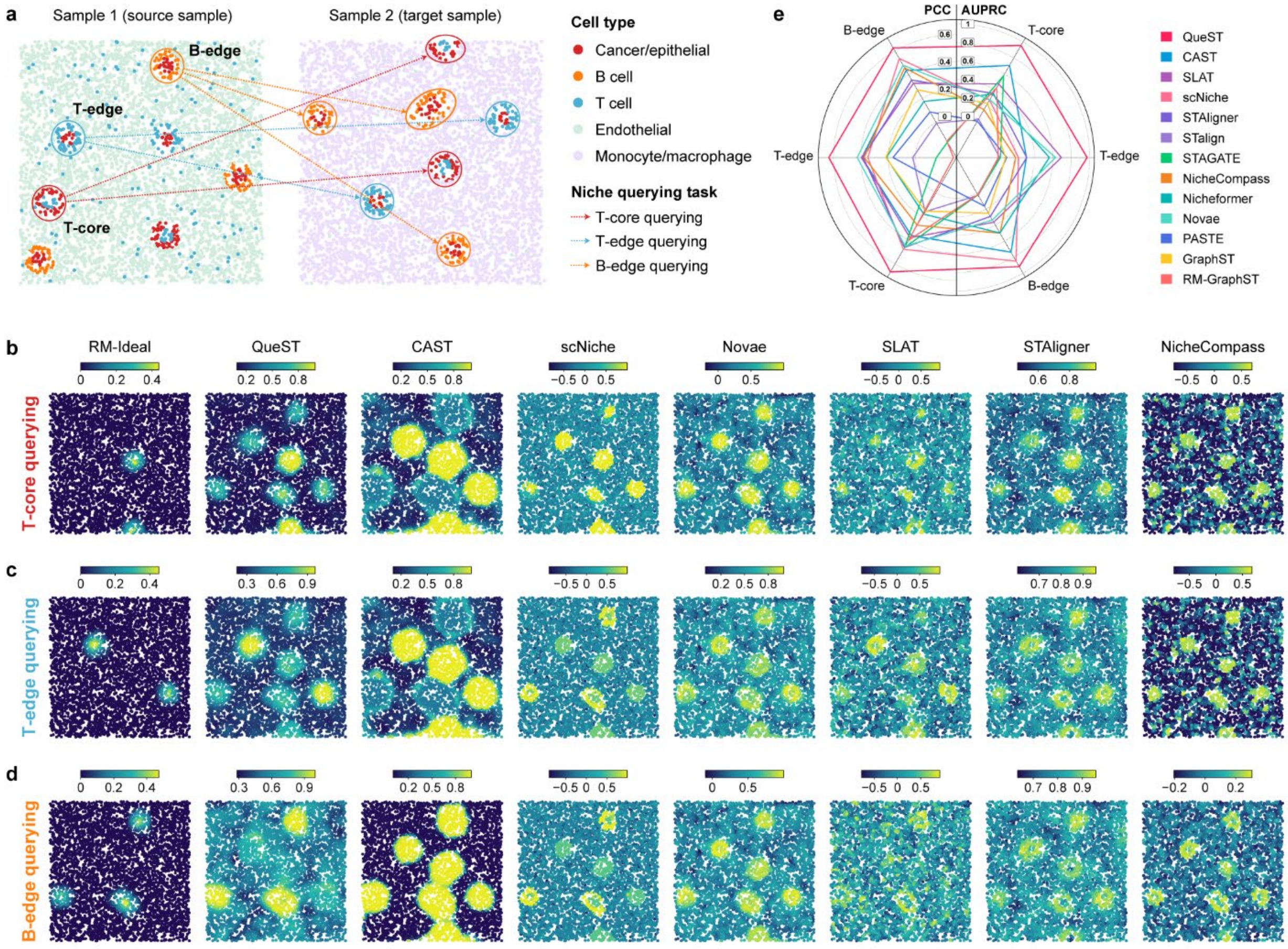

**Fig. 2. Querying niches across topological and environmental variations. a.** Ground truth cell type annotation of the simulated ST samples and schematic illustration of the three niche-query tasks. **b.** Predicted matching scores of different methods for T-core querying, methods with the highest average AUPRC were included. **c.** Predicted matching scores of different methods for T-edge querying, methods with the highest average AUPRC were included. **d.** Predicted matching scores of different methods for B-edge querying, methods with the highest average AUPRC were included. **e.** AUPRC and PCC scores of different methods for T-edge, T-core, and B-edge querying.

## Systematic benchmarking of niche querying on the DLPFC dataset

We next systematically benchmarked niche querying accuracy on real ST data. Since niches with well-established functional annotations remain relatively underexplored in spatial omics, we used structural niches as accessible benchmarking proxies, grounded in the principle that the spatial organization of cells is a key determinant of their biological function. We chose the Human Dorsolateral Prefrontal Cortex (DLPFC) dataset[37], which comprises 12 well-annotated ST samples with a total of 7 layer types. To simulate diverse niche compositions, we defined seven types of NOIs, four involving two layers and three involving three layers, and each niche contains 100 spots (Table S1). For each type of niche, we generated three NOI replicates in each sample, and all query NOIs are visualized in Fig. S3a. We conducted the cross-sample query task for each type of niche, resulting in 2,772 niche-query tasks in total (7 types of niches × 3 replicates × 12 query samples × 11 reference samples, Fig. 3a). Performance was evaluated by computing the PCC between each method's predicted similarity scores and RM-Ideal, which served as a ground-truth reference.

We first assessed performance at the sample level by averaging PCCs across all types of niches for each sample pair. As shown in Fig. S4, QueST demonstrated substantially higher values of approximately 0.6–0.8, while other methods remained largely below 0.5. In addition to higher accuracy, QueST also demonstrated greater stability across sample pairs, indicating robustness to inter-sample variation. We then evaluated performance across different types of niches by aggregating PCCs across all samples. As shown in Fig. 3b, QueST outperformed all other methods, achieving PCC values above 0.7 for five of the seven NOI types, while others mostly remained below 0.5. Importantly, this advantage persisted when NOIs were located within homogeneous regions (Fig. S3b, S5), indicating that the performance gain is not restricted to boundary-defined niches. Sensitivity analyses and ablation studies (Fig. S6, Supplementary Note 4) further showed that QueST is robust to hyperparameter choices (subgraph hop number $K$ and GNN backbone), and identified the contrastive learning module and subgraph-level pooling strategy as the key components driving its performance (Fig. S6c).

To further assess whether the latent space reflects sensible organization with respect to niche composition, we developed a metric named niche composition Jensen-Shannon (NCJS) score (Methods), which quantifies the compositional similarity between niches. For each niche in this dataset, we calculated an NCJS score averaged among its 10 nearest neighbors in the niche embedding latent space. We subsequently averaged this score across all niches to obtain the average NCJS score, which was used to evaluate whether spatial proximity in the embedding space corresponds to compositional similarity (Fig. 3c). QueST achieved average NCJS scores exceeding 0.8, indicating that its latent space captures smooth transitions in layer composition. In contrast, other methods rarely exceeded 0.75, and CAST, SLAT, PASTE, and STalign could not be evaluated due to the absence of a unified latent space across samples (Methods).

As a case study, we analyzed the L5L6 NOI type, which spans between Layer 5 and Layer 6. We visualized niche embeddings from QueST and GraphST using UMAP. We also computed the NCJS score and RM-Ideal score between the query NOI and all candidate niches to measure the similarity of layer composition and niche graph similarity, respectively. In QueST's embedding, both NCJS and RM-Ideal scores formed smooth, localized gradients around the query NOI, indicating that QueST effectively grouped similar niches in its latent space (Fig. 3d). In contrast, GraphST's embeddings showed highly mixed distributions in terms of both metrics (Fig. S7a). While both methods mixed batches well, QueST's latent space preserved the manifold structure, which closely followed the anatomical order of the cortical layers: Layer1 to White Matter (WM). The query NOI was precisely positioned at the Layer 5–6 boundary, consistent with its composition (Fig. 3e and S3a). We can hardly identify such consistency in GraphST's embeddings, with query NOIs surrounded by mixed layer types. We further evaluated the batch integration performance with metrics including AvgBIO, AvgBATCH, and overall scores[38] (Fig. S7b), and QueST consistently showed superior performance on both biological signal preserving and batch removal.

We further visualized the results of different methods on Sample #151673 (Fig. 3f). QueST's predictions closely followed the ground truth defined by RM-Ideal, while the other methods failed to capture this niche-level distinction. GraphST and CAST failed to distinguish correct niches from background regions, and SLAT captured partial signals but produced diffuse and noisy patterns.

In summary, this systematic benchmark on the DLPFC dataset demonstrates that QueST not only achieves higher niche querying accuracy, but also produces robust representations across both sample- and niche-type variations.

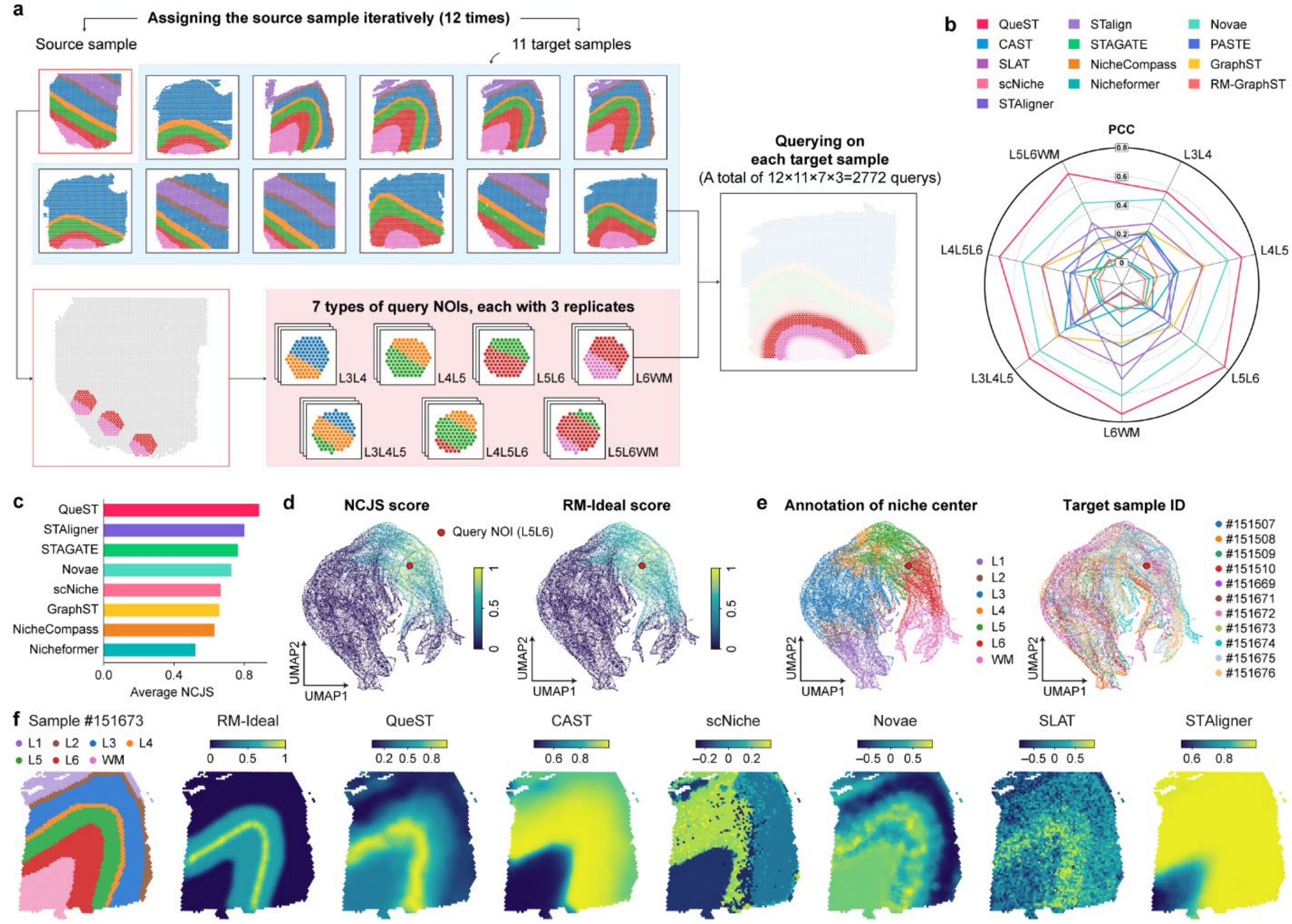

**Fig. 3. Systematic benchmarking of niche querying on the DLPFC dataset. a.** Schematic illustration of the niche querying benchmark on the DLPFC dataset. **b.** Performance of niche querying on the DLPFC dataset in the form of niche-type-wise PCCs. **c.** Quantification of how spatial proximity in the embedding space of different methods corresponds to compositional similarity, measured by the average NCJS score. **d.** UMAP visualization of QueST's niche latent representation colored by the NCJS score and RM-Ideal score. **e.** UMAP visualization of QueST's niche embeddings colored by the layer type annotation of the niche center and sample ID. **f.** Ground truth annotation, RM-Ideal score, and predicted matching scores of different methods for querying L5L6 niche on Sample #151673.

## Querying niches across technologies and spatial resolutions

We next evaluated QueST's ability to identify similar niches across different sequencing technologies and spatial resolutions. We conducted experiments on a Mouse Olfactory Bulb

Tissue dataset[20] (Fig. 4a), a widely studied laminar tissue, which included three samples profiled using 10X Visium (50 μm spot size), Stereo-seq (35 μm), and Slide-seq V2 technologies (10 μm). We set the interface of the granule cell layer (GCL) and mitral cell layer (MCL) as the query NOI, which exhibits a layer-like spatial pattern. We used the Stereo-seq sample as the source sample, while the 10X and Slide-seq V2 samples were target samples.

We computed the PCCs between each method's predicted matching scores and the RM-Ideal score (Fig. 4b). QueST achieved PCCs above 0.8 on both target samples, maintaining consistently the highest performance across both platforms. In contrast, several methods showed platform-dependent behavior. For example, Novae and scNiche achieved reasonable PCC on one platform but notably lower performance on the other. Spatial visualizations revealed that QueST best aligned with RM-Ideal on the 10X sample (Fig. 4c). The other methods all extended the queried region to the tissue margin, especially for NicheCompass and STAGATE, which extended the queried region completely into unrelated layers such as the glomerular layer (GL) and olfactory nerve layer (ONL).

As for the Slide-seq V2 sample, QueST again showed strong agreement with RM-Ideal (Fig. 4d). CAST also produced a comparable spatial pattern, whereas the other methods exhibited diffuse or misplaced predictions. GraphST, STAGATE, and SLAT failed to define a clear queried region, displaying diffuse and ambiguous patterns. NicheCompass, while marking a distinct queried region, identified the wrong region and again extended into GL and ONL.

In summary, QueST consistently outperformed other methods on the Mouse Olfactory Bulb Tissue dataset, achieved high accuracy in localizing the queried region across all platforms, and demonstrated strong cross-platform generalizability. These results demonstrate that QueST can identify shared niche patterns across resolutions without explicit resolution harmonization, relying instead on the consistency of compositional and topological patterns captured by its niche embeddings (Supplementary Note 5).

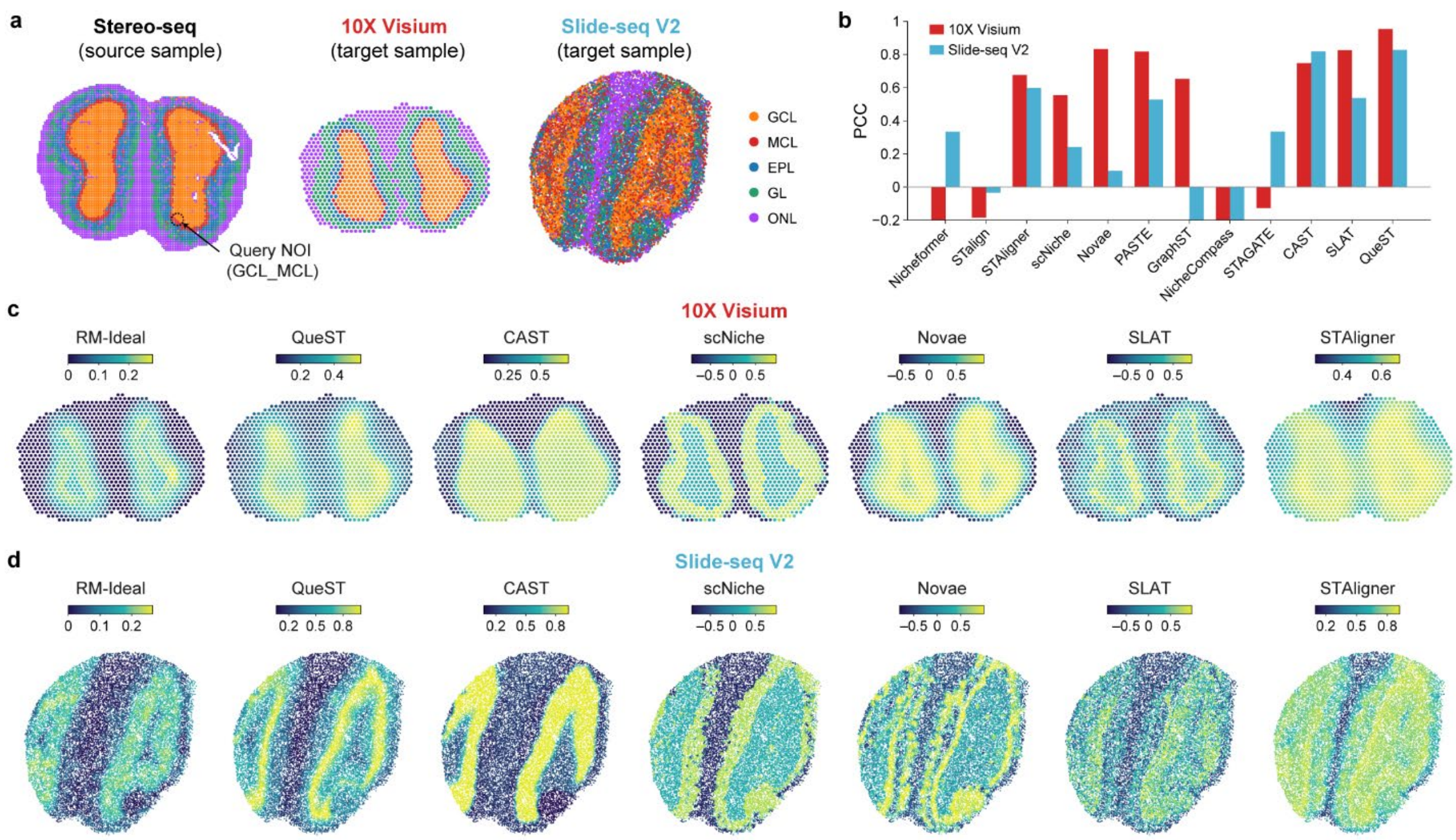

**Fig. 4. Querying niches across different technologies and spatial resolutions with QueST. a.** Ground truth annotation of the Mouse Olfactory Bulb Tissue dataset. **b.** PCCs between the RM-Ideal score and predicted matching scores of different methods. **c.** Spatial distribution of the RM-Ideal score and predicted matching scores of different methods on the 10X sample. **d.** Spatial distribution of the RM-Ideal score and predicted matching scores of different methods on the Slide-seq V2 sample.

## QueST reveals TLS functional heterogeneity for clear cell renal cell carcinoma

Tertiary Lymphoid Structures (TLS) are key niches related to anti-tumor immune responses, commonly observed in tumor microenvironments (TME)[39–41]. While TLS presence is generally associated with favorable anti-tumor immunity, recent studies suggest the existence of functional heterogeneity within TLS, which may underlie divergent and sometimes even adverse clinical outcomes[42–45]. To investigate such heterogeneity, we applied QueST to a clear cell renal cell carcinoma (ccRCC) dataset[46] consisting of 18 TLS-positive ST samples profiled with 10X Visium technology (Fig. S8). We excluded 3 samples (c_2, c_5, and c_23) flagged by the authors for substantially lower transcript capture (Methods). The remaining 15 samples contained 28 spatially separated TLS regions, which were manually annotated by pathologists based on hematoxylin and eosin (H&E) staining. We trained QueST across these samples and obtained TLS embeddings. UMAP visualization of these embeddings indeed revealed substantial heterogeneity among TLSs: we observed two distinct groups of TLSs, denoted as Type A and Type B (Fig. 5a). Noting substantial heterogeneity within Type B, we further subdivided it into two subtypes, B1 and B2.

We compared Type A and Type B using differential expression analysis and identified the top 10 upregulated genes from each type as the signature gene set. To assess their clinical relevance, we evaluated these signatures in an independent ccRCC cohort[47] (JAVELIN 101) treated with immune checkpoint blockade (ICB) therapy (Methods). For each patient, single-sample gene set enrichment scores (ssGSEA) were computed from bulk RNA-seq data for each signature gene set, and patients were stratified into high- and low-score groups for survival analysis. This analysis revealed a significant prognostic divergence: patients with high Type-A-signature scores showed worse outcomes, while patients with high Type-B-signature scores showed better survival (Fig. 5b).

Consistent with their favorable clinical association, Type-B TLSs exhibited elevated expression of multiple immune checkpoint genes, including CD274 (PD-L1), LAG3, HAVCR2 (TIM-3), PDCD1 (PD-1) and TIGIT (Fig. S9a), together with enrichment of immune-related pathways such as antigen processing and presentation, Th1 and Th2 cell differentiation, and IFN-γ response (Fig. 5c). In contrast, Type-A TLSs were characterized by enrichment of pathways related to TGF-β signaling, epithelial–mesenchymal transition (EMT), and ECM–receptor interaction, suggesting a stromal remodeling–dominant state (Fig. 5c).

We further selected TLSs of comparable size and performed cell type deconvolution against a ccRCC scRNA-seq reference dataset[48] using cell2location[49], followed by analysis of radial changes in cell-type composition from the TLS interior to the periphery (Methods). Notably, Type-A TLSs exhibited a higher overall proportion of fibroblasts and epithelial cells, with an increasing gradient from the core to the outer regions, while tumor cell abundance remained relatively low throughout the niche (Fig. 5d–e). Immune populations, such as B cells and $CD4^+$ T cells, were also abundantly present within Type-A TLSs, and their relative abundance gradually decreased from the core toward the periphery. The pathway activity related to TGF-β signaling displayed a similar outward enrichment pattern (Fig. 5d–e), spatially coinciding with the peripheral enrichment of fibroblasts (Fig. 5d–e). These coordinated spatial trends suggest a stromal-dominated niche architecture in Type-A TLSs, in which peripheral accumulation of fibroblasts and associated TGF-β signaling may cooperatively constrain how immune cells infiltrate and are spatially distributed within the TLS, a configuration consistent with stromal-driven immune exclusion and potentially reduced responsiveness to ICB therapy. This interpretation is supported by prior studies showing that fibroblast-associated TGF-β signaling and stromal remodeling can create a physical and functional barrier that limits immune cell infiltration or confines them to specific spatial compartments within the tumor microenvironment[50–53].

We next examined intra-type heterogeneity within Type B TLSs. Survival analysis revealed further prognostic stratification: patients enriched for the Type-B1 signature showed relatively worse outcomes, while those enriched for the Type-B2 signature showed the most favorable survival (Fig. 5b). Immune checkpoint genes were significantly upregulated in Type-B2 relative to Type-B1 across all five examined markers (Fig. S9b). GSEA further revealed enrichment of IFN-α response, IFN-γ response, oxidative phosphorylation, glycolysis, and hypoxia pathways in Type-B2 (Fig. 5c).

We next examined the spatial organization of cells within Type-B TLSs (see details in Methods). Notably, Type-B2 TLSs exhibited a well-defined spatial pattern (Fig. 5d–e), with immune cells, such as B cells and CD8+ T cells, preferentially enriched in the core region, while tumor cells were predominantly localized at the periphery. Consistently, IFN-α and IFN-γ response signals were elevated within the TLS core, whereas hypoxia-related signals were enriched at the outer boundary. In contrast, Type-B1 TLSs lacked such spatial organization of cell types, particularly with macrophages and tumor cells displaying relatively diffuse and intermixed spatial distributions, while both immune and metabolic pathway activities remained relatively uniform and low across the niche.

These observations suggest that Type-B2 TLSs reside at a structured tumor–immune interface, characterized by a hypoxic tumor-adjacent periphery and a relatively oxygen-permissive, immune-active TLS core. This spatial configuration may facilitate effective sensing of tumor-derived signals at the boundary while preserving immune functionality within the core, thereby supporting a pre-existing but functionally constrained anti-tumor immune response that can be effectively reinvigorated by ICB therapy. In contrast, the absence of such spatial stratification in Type-B1 TLSs suggests that these niches may not be positioned at an effective tumor–immune interface, thereby limiting their contribution to anti-tumor immunity. More broadly, these results suggest that effective TLS function depends not only on immune cell abundance but critically on their spatial organization and positioning within the tumor microenvironment.

Collectively, our analyses suggest that effective anti-tumor immunity within TLSs is associated with overcoming two distinct barriers: sufficient immune infiltration past stromal restriction, which distinguishes Type-B from Type-A, and spatially organized tumor–immune engagement, which distinguishes Type-B2 from Type-B1. QueST's niche-level representation enables the dissection of these layered determinants, providing insights into the functional heterogeneity of TLSs within the tumor microenvironment.

To assess whether the TLS subtype distinction is specific to QueST, we applied GraphST, STAGATE, and NicheCompass to the same dataset. None of these methods produced embeddings that clearly separated the three TLS subtypes or revealed alternative clustering structures, and unsupervised clustering on their embeddings yielded no significant survival associations (Fig. S10). These results highlight the effectiveness of QueST in capturing TLS heterogeneity in this setting.

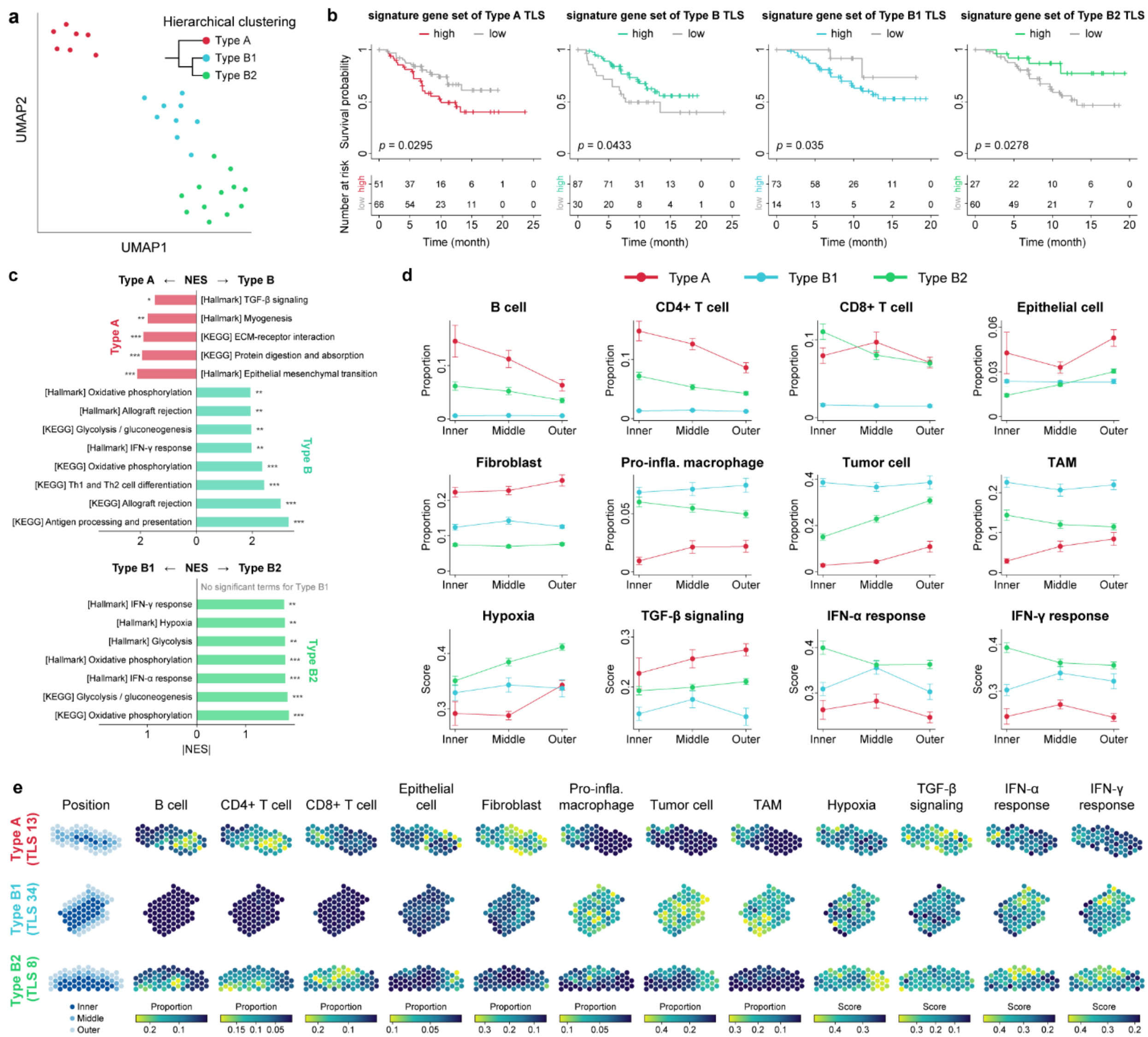

**Fig. 5. QueST reveals TLS functional heterogeneity for clear cell renal cell carcinoma. a.** UMAP visualization of TLS niche embeddings obtained from QueST, colored by TLS subtype (Types A, B1, and B2). **b.** Kaplan-Meier survival curves of overall survival in the JAVELIN 101 cohort, stratified by ssGSEA scores based on the top 10 upregulated genes of each TLS subtype. *P*-values were determined by restricted mean survival time (RMST) analysis with a 15-month truncation time. **c.** GSEA results of Hallmark/KEGG pathways enriched in differentially expressed genes (DEGs) of Type-A vs Type-B (top) and Type-B1 vs Type-B2 (bottom). No pathways were significantly enriched in Type-B1 at FDR < 0.05. Asterisks indicate FDR significance levels: *: FDR < 0.05; **: FDR < 0.01; ***: FDR < 0.001. **d.** Radial distributions of cell-type proportions and gene set/pathway activity scores across different TLS regions. TLS

with 50–100 spots were considered. Dots indicate the mean, and error bars indicate the standard error of the mean (SEM). **e**. Spatial distributions of cell-type proportions and gene set/pathway activity scores in representative TLSs.

## QueST reveals TLS structural heterogeneity across cancer types

Beyond intra-cancer analyses, QueST is also well-suited for cross-cancer comparisons of spatial niches, enabling the identification of shared and distinct structural features across tumor types. Increasing evidence suggests that multicellular spatial communities such as TLSs may function as reusable tissue modules rather than strictly cancer-specific structures[15,54–56]. Within this context, cross-cancer niche mapping provides a direct way to test whether spatial architectures learned in one tumor type can be transferred to another, thereby probing the degree of structural conservation versus context specificity, and offering insights into immune regulation and therapeutic response mechanisms[10,57,58].

To evaluate this capability, we applied QueST to map the aforementioned ccRCC TLSs onto a non-small cell lung cancer (NSCLC) dataset[59]. This dataset contained three ST samples with cell type and niche annotations (Fig. S11), profiled using the imaging-based single-cell resolution Nanostring CosMx SMI technology, with a panel of 960 genes. We performed the mapping of TLS by using TLS niches from ccRCC as query NOIs to retrieve matching candidate niches in NSCLC samples. Considering the technological and biological gap between these two datasets, we first validated the accuracy of TLS mapping. The accuracy was evaluated using the area under the receiver operating characteristic curve (AUROC), based on ground-truth annotations in the NSCLC samples (Fig. 6a). QueST consistently achieved AUROC scores above 0.9 across all samples, demonstrating superior accuracy. In contrast, other methods encountered great challenges and produced inferior results. For example, STAGATE barely exceeded an AUROC of 0.7. CAST, while achieving an AUROC near 0.9 on one sample, dropped to ~0.6 on Sample-3, indicating low robustness.

To avoid potential evaluation bias caused by large TLS regions dominating the results, we visualized the spatial distribution of ground truth annotation and TLS query scores of different methods (Fig. 6b, Fig. S12). QueST accurately identified TLS regions regardless of the area size, demonstrating its robustness against the size variations among TLS regions. In contrast, other methods produced diffuse patterns and failed to clearly separate TLS from surrounding tissue.

We next examined how TLS subtypes from ccRCC (Types A, B1, and B2) are represented in NSCLC. We performed subtype assignment within QueST-identified TLS regions in NSCLC (Methods). This analysis revealed a skewed subtype composition (Fig. 6c), with TLS niches distributed between Type-A and Type-B1, while no niches were assigned to Type-B2, suggesting that the highly organized, immune-activated TLS state represented by Type-B2 in ccRCC is not present in these NSCLC samples.

We further characterized the structural and cellular differences between the queried Type-A and Type-B1 niches. Spatially, Type-A niches were preferentially located adjacent to tumor regions, whereas Type-B1 niches were positioned farther from the tumor boundary, in line with the notion that Type-B1 TLSs are not situated at an effective tumor–immune interface (Fig. 6c, Fig. 6d). Consistent with this tumor-proximal localization, epithelial cells were detected in Type-A niches but were nearly absent from Type-B1 niches (Fig. 6e), further supporting stronger tumor–TLS coupling in Type-A.

We also observed distinct stromal organization patterns. Though the overall abundance of fibroblasts in Type-A niches was slightly lower than in Type-B1 (Fig. 6e), fibroblasts in Type-A niches were preferentially enriched at the TLS periphery, recapitulating the spatial distribution pattern observed in ccRCC (Fig. 6f). Further analysis revealed that these fibroblasts exhibited upregulation of canonical cancer-associated fibroblast (CAF) programs, as well as EMT and ECM–receptor interaction gene sets, suggesting a stromal remodeling–dominant microenvironment (Fig. 6g).

In summary, QueST enables accurate and robust cross-cancer TLS mapping and provides a framework for dissecting TLS heterogeneity through query-based niche stratification. This analysis reveals that while certain structural features of TLSs are conserved across cancer types, their spatial context and stromal coupling can vary substantially, potentially underlying functional divergence among TLS subtypes.

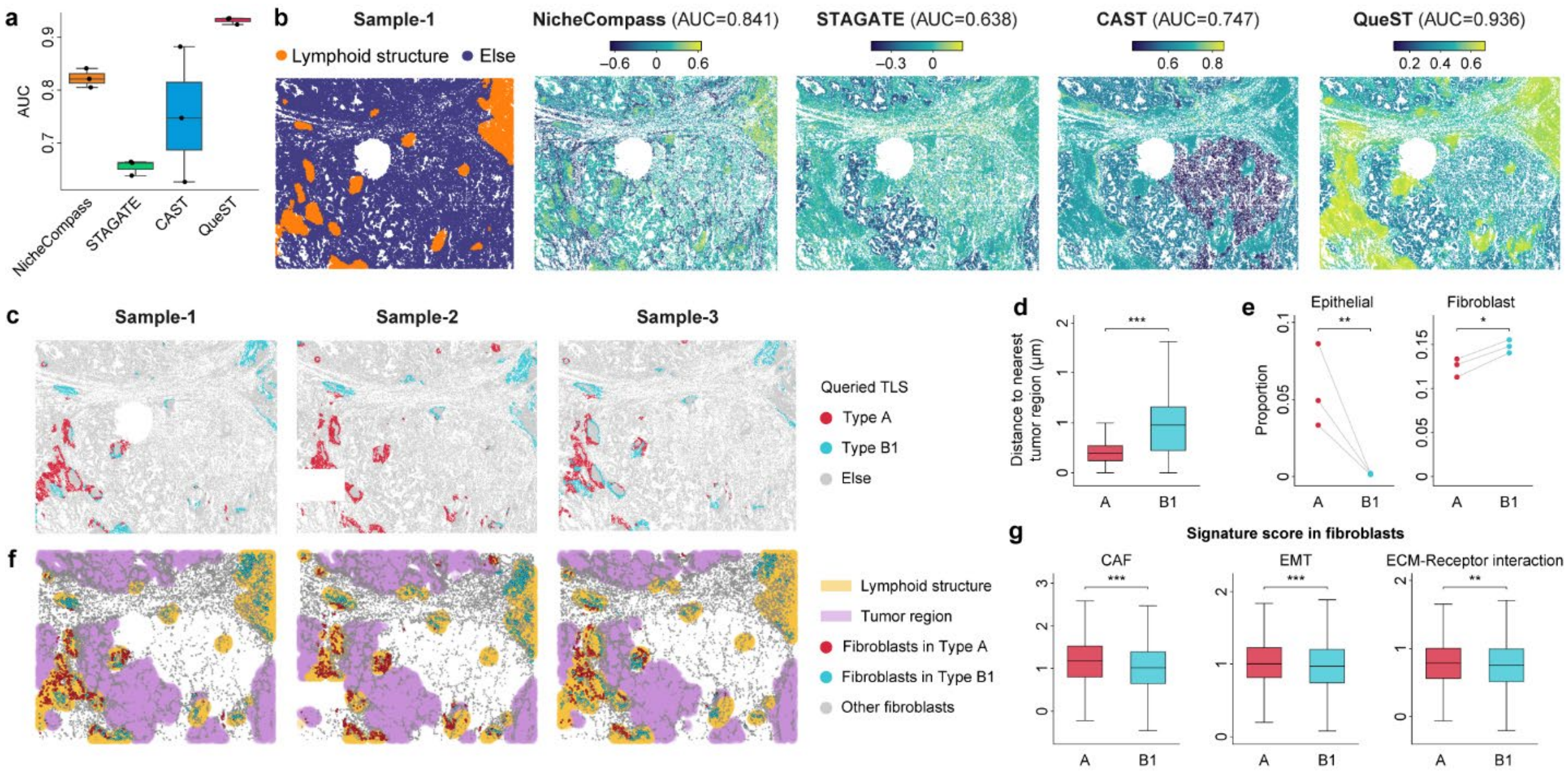

**Fig. 6. QueST reveals TLS structural heterogeneity via querying across cancer types. a.** Boxplot showing the AUROC scores of different methods for TLS detection on NSCLC samples. **b.** Ground truth annotation and spatial distribution of predicted matching scores of different methods on NSCLC Sample-1. **c.** Spatial distribution of queried TLS regions identified by QueST on NSCLC samples. **d.** Spatial distance between queried TLS regions and tumor regions. **e.** Per-sample epithelial and fibroblast cell fractions in Type-A versus Type-B1 queried regions. Each dot represents the corresponding cell-type fraction within queried TLS regions from one NSCLC sample. *P*-values were calculated using paired *t*-tests on logit-transformed fractions. **f.** Spatial distribution of fibroblasts colored by queried TLS regions types (A-fibro, B1-fibro, other) overlaid on annotated lymphoid structure and tumor regions in the three NSCLC samples. **g.** CAF, EMT (Hallmark), and ECM–receptor interaction (KEGG) signature scores in fibroblasts from Type-A and Type-B1 TLSs. *P*-values were calculated by two-sided Mann–Whitney *U* tests. Asterisks indicate significance levels: *: < 0.05; **: < 0.01; ***: < 0.001.

## *De novo* NOI discovery and cross-sample retrieval using QueST

In the preceding analyses, NOIs were defined based on prior biological knowledge, such as pathologist-annotated TLS regions. However, in exploratory settings, researchers encounter spatial transcriptomics datasets without knowing what functionally relevant niches exist or where they are located. Here, we demonstrate that QueST supports a complete discovery-oriented workflow, from *de novo* NOI discovery to cross-sample niche retrieval and characterization.

We applied QueST to an *in vivo* combinatorial perturbation ST dataset[60] comprising six tissue sections from a mosaic liver cancer mouse model containing 324 tumor nodules, each harboring distinct combinations of eight oncogenic perturbations. We trained QueST across all sections, obtained niche embeddings (Fig. 7a), and selected sample ML_III_B as the source sample due to its highest nodule density. To prioritize biologically interesting NOI candidates, we defined a niche prevalence ratio (NPR) score to quantify how prevalent a niche pattern is across samples relative to its proportion within the source sample (Methods), and selected the highest-NPR niche in the source sample as NOI (Fig. 7b).

The NOI localized to the core of a tumor nodule annotated as Hamp2/Upp2 type in the original study (Fig. 7b). We retrieved niches sharing the NOI's spatial pattern across all reference samples and found that the queried regions consistently localized to nodule cores (Fig. 7c). Notably, in sample ML_II_B, the queried region precisely overlapped with the core of nodule N42, which had been labeled as "unclassified" in the original study (Fig. 7c). To assess its molecular identity, we examined the relative expression of Hamp2/Upp2 marker genes within the queried region, following the criteria defined in the original study, where a nodule is classified as Hamp2/Upp2 type if at least two of four marker genes (Car3, Upp2, Hamp, and Hamp2) exceed a predefined threshold (Methods). The retrieved core region of N42 satisfied this criterion, with two marker genes exceeding the threshold (Fig. 7d). Notably, this classification was not assigned in the original study, likely because the original annotation was performed over the entire nodule, diluting the marker signal from the core region. In contrast, QueST focuses on functionally coherent subregions, enabling the recovery of molecular identities that may be obscured at the whole-nodule level.

We next extended this analysis to all retrieved nodules in the target samples. The queried regions consistently exhibited higher Hamp2/Upp2 marker gene expression than the corresponding nodules as a whole (Fig. 7e), indicating that QueST preferentially retrieved subregions with molecular characteristics consistent with the NOI. Applying the same classification criteria, 9 of the 14 retrieved nodules were classified as Hamp2/Upp2 type based on the queried regions. For the remaining 5 nodules that did not meet the threshold (Fig. 7f), their queried regions nevertheless showed the highest Hamp2/Upp2 marker expression compared to all other nodule types defined in the original study (Fig. 7g), suggesting that they represent closely related or partial variants of this category. Consistently, analysis of perturbation composition revealed that all retrieved nodules carried the NICD perturbation with high probability (Fig. 7h), in agreement with its reported association with the Hamp2/Upp2 type[60]. These results support the molecular coherence of the retrieved niches.

Collectively, these results demonstrate that QueST supports a *de novo* discovery-oriented workflow, from data-driven NOI discovery to cross-sample querying and characterization, without requiring prior biological annotation. This extends QueST beyond querying predefined structures and supports its use as a general framework for uncovering previously unannotated functional niches.

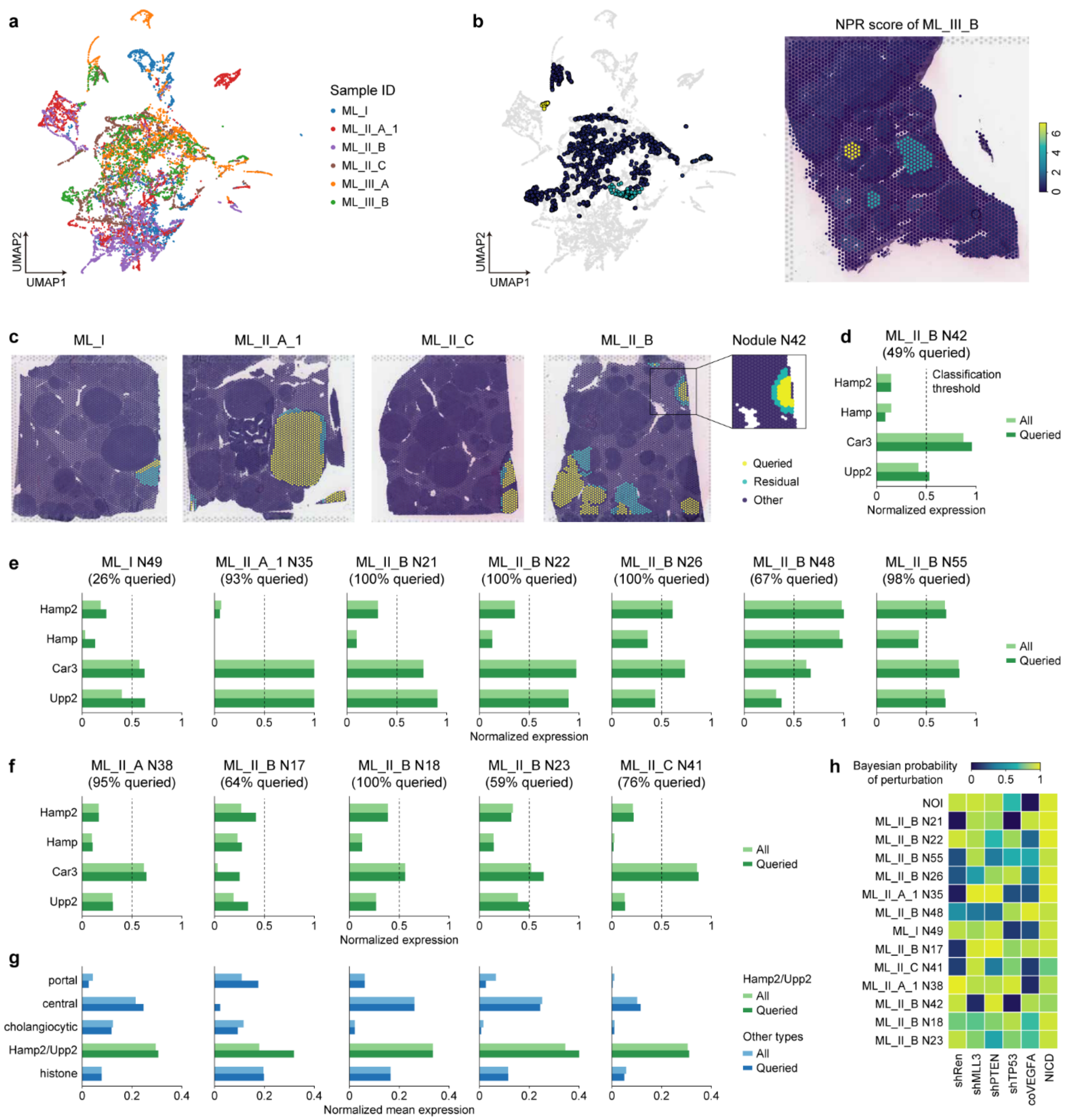

**Fig. 7. *De novo* NOI discovery and cross-sample querying on a combinatorial perturbation spatial transcriptomics dataset. a.** UMAP visualization of QueST's global niche embeddings across six tissue sections, colored by sample ID. **b.** Distribution of NPR scores on the global UMAP and the source sample ML_III_B, with the highest-scoring region corresponding to the selected NOI. **c.** Spatial mapping of the queried niche pattern across reference samples. Yellow regions denote queried niches; blue regions denote the residual parts of corresponding nodules; dark regions denote other tissue areas. **d**. Expression levels of the four Hamp2/Upp2 marker genes in the queried region compared with the entire nodule N42 from sample ML_II_B. **e**.

Expression levels of the four Hamp2/Upp2 marker genes across all retrieved nodules whose queried regions exceeded the classification threshold, comparing queried regions with the corresponding entire nodules. **f**. Expression levels of the four Hamp2/Upp2 marker genes across retrieved nodules whose queried regions remained below the classification threshold. For (d–f), dashed lines indicate the classification threshold defined in the original study, which requires at least two Hamp2/Upp2 marker genes to exceed the threshold. Percentages denote the proportion of spots within each nodule that were included in the queried region. **g.** Mean expression level of marker genes for five nodule types in the unclassified nodules, comparing all versus queried regions. **h.** Perturbation composition of all queried nodules. Color intensity indicates the Bayesian probability of each perturbation existing within each nodule. Two global perturbations (Myc, Mctnnb1) were filtered due to a lack of nodule selectivity.

## Discussion

In this study, we defined the niche-query task and developed QueST, a powerful method for querying spatial niches across diverse ST samples from both sequencing-based (such as 10X Visium) and imaging-based (such as Nanostring CosMx SMI technology) spatially resolved transcriptomics platforms. QueST offers an effective way of learning niche representations via subgraph-based modelling and contrastive learning, and enables quantitative similarity comparison and querying of niches based on the similarity. Experiments have shown the success of QueST in learning the gene expression and structural information of niches across platforms with diverse gene detection sensitivities, and the application of cross-sample TLS mapping and niche discovery in perturbation settings has revealed the value of the niche-query task for unveiling both shared and distinct organizational patterns across biological conditions.

With the rapid development of ST technologies and the growing availability of ST datasets, an increasing number of spatial niches with distinct functions and diagnostic potential are being uncovered. However, systematically identifying niches that resemble known functional niches across tissues and individuals remains an open challenge. We anticipate that niche querying will become an essential component of spatial omics analysis, supporting comparative studies of tissue organization and disease microenvironments.

By learning informative and robust representations of spatial niches, QueST provides a quantitative and biologically meaningful metric of niche similarity. This allows not only targeted querying, but also lays the foundation for atlas-scale niche comparison, integration, and classification. Just as single-cell transcriptomics has led to the establishment of a unified cell type taxonomy[61,62], and cross-tissue coordinated cellular modules[15], we believe that the next frontier is to define and organize niche types as recurrent multicellular units with conserved spatial architectures and biological functions, which will advance our understanding of tissue organization beyond the single-cell level.

The design of QueST defines spatial niches using fixed-radius $K$-hop subgraphs and learns representations within this local context. This design is largely resolution-agnostic in practice, as it does not rely on explicit resolution harmonization but instead applies a uniform subgraph-based representation across datasets, regardless of the sequencing platform or spatial resolution. Given an NOI, QueST retrieves similar regions by comparing embeddings of local subgraphs, thereby operating at the level of compositional and topological patterns rather than absolute spatial scale. This works in practice because, at the level of local niches, biologically similar structures tend to have consistent patterns of cell-type composition and relative spatial organization, even when measured at different resolutions. As a result, subgraphs corresponding to analogous regions (e.g., boundary-like or core-like structures) can produce similar embeddings despite differences in physical scale, as demonstrated by our cross-platform experiments. This robustness extends to scenarios with differences in both resolutions and gene detection sensitivity. In our cross-cancer TLS analysis, despite a >10-fold disparity in gene coverage between the source (10X Visium, ~17,000–36,000 genes) and target (Nanostring CosMx SMI, 960 genes) samples, QueST achieved AUROC values above 0.9 for TLS retrieval across all three NSCLC samples. However, in cases of extreme resolution disparity or niches with highly heterogeneous internal structures, this pattern-level correspondence may become insufficient to fully capture fine-grained spatial organization. In such scenarios, we represent higher-order structures such as TLSs as aggregations of multiple local subgraph embeddings of QueST, effectively modeling each niche as a composite of sub-niches capturing different aspects of its internal organization. This pooling-based design enables meaningful comparison across resolutions by integrating heterogeneous local patterns into a unified representation. However, this strategy does not explicitly model spatial relationships between sub-niches, such as their relative arrangement or higher-order topology, and may therefore lose certain structural information.

A promising future direction is to extend this formulation towards multi-scale and hierarchical[63] modeling of spatial structures, which would allow niche representations to adapt to the native resolution of each dataset rather than relying on a fixed *K*. Such extensions could represent niches as structured collections of subgraphs and explicitly model relationships across scales, enabling more faithful representation of complex spatial organization. Moreover, as ST datasets continue to grow in scale and diversity, adapting QueST to more scalable architectures will be important. Inspired by recent progress in foundation models for single-cell and spatial transcriptomics[64–71], a natural extension of QueST would be to develop a foundation model tailored for learning representations for niches on large-scale spatial transcriptomics data.

# Methods

## Model architecture of QueST

The QueST model has three modules: the asymmetric graph auto-encoder (GAE) module that learns node embeddings for individual cells from the gene expression data and spatial proximity graph, the contrastive learning (CL) module that aggregates node embeddings into niche embeddings and refines them via subgraph-level contrastive learning, and the adversarial batch removal module that applies adversarial training to ensure batch-invariant niche representations.

Each ST sample is modeled as a spatial proximity graph, with each cell or spot as a node and the gene expression as the initial node features. QueST defines each candidate niche as a $K$-hop subgraph centered at a node, and learns a corresponding niche embedding. These embeddings are used for downstream niche querying across samples.

### Construction of the spatial proximity graph

Spatial proximity graphs are constructed using the spatial coordinates of cells, with each sample represented as an undirected graph $G = (V, E, X)$, where each node $v \in V$ corresponds to a cell, $E$ denotes the edge set, and $X$ is the node feature matrix derived from gene expression. We use the `squidpy.gr.spatial`[72] function to construct this graph, where we choose the `coord_type` parameter according to the spatial coordinate layout. Specifically, for grid-like spatial coordinate layout, such as samples generated via 10X Visium or Stereo-seq sequencing technology, we set `coord_type` as "`grid`", while for the other single-cell resolution datasets, we set `coord_type` as "`generic`", and the spatial graph was determined using Delaunay triangulation.

**Graph Auto-encoder**

The graph auto-encoder module takes normalized gene expression as input and reconstructs it as the output. It is built upon a modified Graph Isomorphism Network (GIN)[73] layers, where we added residual connections between conventional GIN layers to enhance model robustness. The auto-encoder is asymmetric, consisting of a three-layer GIN encoder and a single-layer decoder. In each GIN layer, the node embedding of node $v$ is updated via aggregating all the messages from its neighbors:

$$\boldsymbol{h}_v^{(l+1)} = \mathrm{MLP}^{(l)}\left((1+\epsilon)\cdot\boldsymbol{h}_v^{(l)} + \sum_{u\in\mathcal{N}(v)}\boldsymbol{h}_v^{(l)}\right) + W^{(l)}\boldsymbol{h}_v^{(l)} \tag{1}$$

where $\boldsymbol{h}_v^{(l)}$ is the node embedding of node $v$ in the $l$-th GIN layer, and $\boldsymbol{h}_v^{(\mathbf{0})} = \boldsymbol{x}_v$ which is the input normalized gene expression vector. $\mathcal{N}(v)$ is the node set includes all neighbor nodes of $v$, and $\mathrm{MLP}^{(l)}$ was a 2-layer fully connected neural network serving as the non-linear projection function of the GIN layer. $\epsilon$ is a scalar that controls the weight of each node's own feature in the aggregation process, which is set to 0 for all experiments in this study. $W^{(l)}$ is a learnable linear transformation added for ensuring dimension consistency in residual connection.

The encoding and decoding process of the graph auto-encoder can be described as follows:

$$\boldsymbol{z}_v = f_{enc}(\boldsymbol{x}_v) = \boldsymbol{h}_v^{(l_{enc})} \tag{2}$$

$$\widehat{\boldsymbol{x}}_v = f_{dec}(\boldsymbol{z}_v) = \boldsymbol{h}_v^{(l_{enc}+l_{dec})} \tag{3}$$

where $\boldsymbol{z}_v$ is the latent node embedding of node $v$, $l_{enc}$ and $l_{dec}$ are the number of GIN layers in the encoder and decoder respectively, and $\widehat{\boldsymbol{x}}_v$ is the reconstructed normalized gene expression vector. The loss function for reconstruction is the Mean Squared Error (MSE):

$$L_{\text{reconstruction}} = \frac{1}{|V|}\sum_{v\in V}\left|\left|\hat{\boldsymbol{x}}_v - \boldsymbol{x}_v\right|\right|_2 \tag{4}$$

**Contrastive niche representation learning**

To obtain high-quality niche embeddings that encode both functional and structural information, we developed a subgraph-based CL module. This module takes node embeddings and the spatial graph as input and outputs structure-aware niche embeddings. Specifically, we first defined the $K$-hop subgraph of each node as the niche centered at this node:

$$\text{niche}_v = \text{k_hop_subgraph}(V, E, v) \tag{5}$$

where $\text{niche}_v$ is the set of nodes within $K$-hop distance from node $v$, including $v$ itself. We set $K = 3$ in all experiments, motivated by prior work on subgraph-based modeling of cellular microenvironments[30] (see Supplementary Note 4 for details).

The initial niche embedding $\boldsymbol{s}_v$ is computed by mean pooling the node embeddings within the subgraph:

$$\boldsymbol{s}_v = \frac{1}{|\text{niche}_v|}\sum_{u\in \text{niche}_v} \boldsymbol{z}_u \tag{6}$$

where $\boldsymbol{s}_v$ is the niche embedding for the aforementioned niche. Simply pooling over node embeddings may overlook topological structures and lead to over-smoothing. To overcome this challenge, we designed a subsequent contrastive learning step to refine the niche embeddings, and the detailed procedure is described as follows:

**Node fixing**: A certain subset of nodes, denoted as $V_{\text{positive}}$, is randomly sampled. For each node in $V_{\text{positive}}$, we denote its corresponding niche as a "fixed niche", which implies all nodes in its corresponding niche are fixed during data corruption. The fixed node set is:

$$V_{\text{fix}} = \{u \in \text{niche}_v | v \in V_{\text{positive}}\} \tag{7}$$

**Data corruption**: A corrupted graph $G' = \left(V, E, X'\right)$ is then generated by randomly shuffling features (gene expression) among unfixed nodes $V_{\text{negative}} = V \setminus V_{\text{fix}}$, while preserving the original graph structure. This yields partial feature corruption without disturbing the spatial structure of selected niches. We denote a niche with its center node in $V_{\text{negative}}$ as a "shuffled niche", and used those with shuffled proportion in the range of $[\alpha, \beta]$ as negative samples in the CL module. We set $\alpha = 0.25, \beta = 0.75$ in all experiments.

**Dual encoding**: Both the original graph $G$ and the corrupted graph $G^{'}$ are then encoded by the same graph encoder. For each niche $\text{niche}_v$ centered at $v \in V_{\text{positive}}$, we obtain a pair of niche embeddings $\boldsymbol{s}_v$ and $\boldsymbol{s}_v^{'}$, representing the same niche under original and corrupted conditions. The positive pairs and negative pairs are defined as $\left(\boldsymbol{s}_v,\ \boldsymbol{s}_v^{'}\right)$ for $v \in V_{\text{positive}}$, and $\left(\boldsymbol{s}_v,\ \boldsymbol{s}_u^{'}\right)$ for $u \in V_{\text{negative}}$, respectively.

**Contrastive discriminating**: A bilinear discriminator $g_{\text{contrast}}: \mathbb{R}^d \times \mathbb{R}^d \to \mathbb{R}$, following Deep Graph Infomax[74], is used to score similarity between niche pairs, where $d$ is the dimension of QueST's niche embeddings. The contrastive loss is:

$$L_{\text{contrast}} = -\frac{1}{\left|V_{\text{positive}}\right|} \sum_{v \in V_{\text{positive}}} \left[\log\left(g_{\text{contrast}}\left(\boldsymbol{s}_v, \boldsymbol{s}_v^{'}\right)\right) + \log\left(1 - g_{\text{contrast}}\left(\boldsymbol{s}_v, \boldsymbol{s}_u^{'}\right)\right)\right] \quad (8)$$

This process encourages the model to align embeddings of structurally and transcriptionally similar niches, while distinguishing dissimilar ones.

**Adversarial learning for batch effect removal**

To mitigate batch effect and enable cross-sample comparisons of the niche embeddings, we borrowed the idea of Generative Adversarial Networks[35] (GAN) and developed an adversarial training module. This module introduces a discriminator that takes niche embeddings as input and predicts their batch origin. The encoder is trained against this discriminator in order to generate batch-free niche embeddings. Specifically, we trained QueST jointly on all samples in a merged fashion, and for each sample in each epoch, we have a discrimination step and a generation step:

**Discrimination step**: In this step, a three-layer MLP is used as the batch discriminator, denoted as $g_{\text{batch}}: \mathbb{R}^d \to \mathbb{R}^B$, where $B$ is the number of batches (one per sample in our setting). Given a niche embedding $\boldsymbol{s}_v$ with ground-truth batch ID $c$, the discriminator is trained using the cross-

entropy loss function:

$$L_{\text{batch_discriminator}} = -\frac{1}{|V|}\sum_{v \in V} \log \frac{\exp\left(g_{\text{batch}}^{(c)}(\boldsymbol{s}_v)\right)}{\sum_{j=1}^{B} \exp\left(g_{\text{batch}}^{(j)}(\boldsymbol{s}_v)\right)} \quad (9)$$

Here, $g_{\text{batch}}^{(j)}(\boldsymbol{s}_v)$ is the predicted probability of $\boldsymbol{s}_v$ originating from the $j$-th batch. In this step, the parameters in the encoder are fixed and only the batch discriminator is trained.

**Generation step**: The optimization objective is then reversed. The batch discriminator is frozen, and the encoder is updated to minimize the negative of the discriminator loss:

$$L_{\text{batch generator}} = -L_{\text{batch discriminator}} \quad (10)$$

This reversal drives the encoder to produce niche embeddings that obscure batch identity, thereby encouraging representations that are invariant to batch-specific variation.

**Overall loss function**

During the training process, all three modules of QueST are trained simultaneously. In each training epoch, QueST is trained on all ST samples in shuffled order. For each sample, the loss function of the QueST model in the discrimination step is:

$$L_{\text{discrmination}} = L_{\text{batch_discriminator}} \quad (11)$$

and the loss function in the generation step is:

$$L_{\text{generation}} = L_{\text{reconstruction}} + L_{\text{contrast}} + L_{\text{batch_generator}} \quad (12)$$

**Querying niches in the latent space**

Once the QueST model is trained, the predicted matching score between two niches is computed using the cosine similarity between their corresponding embeddings:

$$\text{matching_score}_{\text{QueST}}(\text{Niche}_u, \text{Niche}_v) = \frac{\boldsymbol{s}_u^{\top}\boldsymbol{s}_v}{\|\boldsymbol{s}_u\|\|\boldsymbol{s}_v\|} \quad (13)$$

For query NOIs that are not in a strict *K*-hop subgraph form (e.g., manually defined or irregularly shaped regions), mean pooling is directly applied over the node embeddings within the specified region to derive the query NOI embedding. The query NOI embedding is subsequently used for computing the matching score.

## Generation of simulation data

Data was generated following the tutorial of scCube https://github.com/ZJUFanLab/scCube/blob/main/tutorial/customized_complex_patterns.ipynb. For the query sample (Sample-1), we set Endothelial as the background cell type, with 2.5% infiltration of T-cells. For the reference sample (Sample-2), we set Monocytes/Macrophages as the background cell type, with 2.5% infiltration of Endothelial. For both samples, we set the total cell number to 5000. We used the `generate_pattern_custom_ring` function to generate the ring shape of different niches, and the `ring_purity` was set to 1. Since scCube directly generated normalized expression data with 2000 genes, we did not perform further normalization for the simulation data.

## Preprocessing of real ST data

**Human dorsolateral prefrontal cortex (DLPFC) 10X Visium dataset**. All 12 samples in this study were used. Each sample contained around 3000-4000 spots and 33538 genes. Spots annotated as NA were excluded. For each sample, we extracted the gene expression matrix, spatial coordinates, and cell type annotations, and saved them in `.h5ad` format. We selected 3,000 highly variable genes per sample and computed the union of gene sets across all samples. Expression data were then normalized and log-transformed using the Scanpy[75] Python package.

**Mouse Olfactory Bulb Tissue (MOBT) dataset**. The 10X Visium sample, Stereo-seq sample, and the Slide-seq V2 sample contained 1185, 8827, and 18537 spots, respectively, with 5531 genes after intersection. The same preprocessing steps were followed: selection of 3,000 highly variable genes per sample, gene set union across samples, and subsequent normalization and log transformation using Scanpy.

**Clear cell renal cell carcinoma (ccRCC) dataset**. Gene expression, spatial coordinates, and TLS annotations were extracted from raw data and saved in `.h5ad` format. Samples without pathologist-annotated TLS regions were excluded, yielding 18 TLS-positive samples (10 FFPE and 8 fresh-frozen), each containing approximately 1,000–5,000 spots. We further excluded 3 frozen samples (c_2, c_5, and c_23) flagged by the authors for substantially lower transcript capture (median genes per spot of 875–2,078, compared with ~3,700 for the remaining samples); c_5 was additionally noted as having poor RNA quality and atypical lobular tumor organization. The final dataset comprised 15 samples (10 FFPE and 5 fresh-frozen) with 28 spatially separated, pathologist-annotated TLS regions. FFPE samples contained 17943 genes, and Frozen samples contained 36601 genes. For preprocessing, we first performed gene intersection across all samples, then selected 3,000 highly variable genes per sample, took the union across samples, and applied normalization and log transformation using Scanpy.

**Non-small cell lung cancer (NSCLC) dataset**. Three samples (Sample 1–3) in the NSCLC dataset were included, with each sample containing around 90000 cells and 960 genes. Prior to cross-cancer comparison with the ccRCC dataset, gene intersection was performed to obtain a shared gene set. No additional gene filtering was applied. The resulting expression matrices were then normalized and log-transformed using Scanpy.

**Perturb-CAST dataset**. Six 10X Visium tissue sections (ML_I, ML_II_A_1, ML_II_B, ML_II_C, ML_III_A, ML_III_B) from a mosaic liver cancer mouse model were used. Each section contained 2,697–4,217 spots and 13,650–14,545 genes. These sections were derived from two mice injected with eight cancer-driving perturbation plasmids via hydrodynamic tail-vein injection, harboring 324 genetically heterogeneous tumor nodules with distinct perturbation combinations. Only tumor spots (those annotated within histologically identified nodules with at least 3 spots) were retained for QueST training, resulting in 13,603 tumor spots across the six sections. For preprocessing, we performed gene intersection across all six samples, selected 4,000 highly variable genes per sample, took the union across samples, and applied normalization and log transformation using Scanpy. Bayesian perturbation integration probabilities and tumor phenotype annotations were obtained from the original study and mapped to individual spots via nodule-level metadata.

## Baseline methods for comparison

We compared QueST against 11 existing methods: GraphST (v1.0.0), STAGATE (v1.0.1), NicheCompass (v0.2.2), CAST (v0.4.0), SLAT (v0.3.0), Nicheformer (v0.0.1), Novae (v1.0.2), PASTE (v1.4.0), scNiche (v1.1.1), STalign (v1.0), and STAligner (v1.0.0). As required by these methods, gene expression data and spatial proximity graphs were used as input. We followed the data preprocessing pipeline recommended by the original authors.

**GraphST**. We trained GraphST on all samples in a single run. We treated the latent node embeddings as niche embeddings. For RM-GraphST in the simulation study, we performed Kmeans clustering using `sklearn.cluster.KMeans` with the number of clusters manually set as the number of types of niches. We then applied the RM method to the cluster labels to obtain the predicted matching score.

**STAGATE**. We trained STAGATE on all samples in a single run. Node embeddings were treated as niche embeddings.

**NicheCompass**. We trained NicheCompass on all samples in a single run. Required gene annotations and cell–cell interaction files were downloaded following the official tutorial. These, along with gene expression and spatial graphs, were used as input. We treated node embeddings as niche embeddings.

**CAST**. We trained CAST on a sample pair in one training run, and iterated over all sample pairs for niche querying across all samples. We obtained the node embeddings from the CAST Mark module as niche embeddings. We obtained the spatial alignment results from the CAST Stack module.

**SLAT**. We trained SLAT on a sample pair in one training run, and iterated over all sample pairs for niche querying across all samples. We extracted the node embeddings from SLAT's LGCN encoder as niche embeddings. We obtained the spatial alignment results from SLAT's `spatial_match` function.

**Nicheformer**. We fine-tuned the pre-trained model on each dataset using domain adaptation with a learning rate of 1e-5 for 5 epochs. For datasets where gene names were not in Ensembl ID format, we mapped gene symbols to human Ensembl IDs using the MyGeneInfo API. We extracted cell embeddings from the last hidden layer and applied mean pooling.

**Novae**. We used the pre-trained Novae foundation model from HuggingFace and fine-tuned it on each dataset with a learning rate of 5e-4 for up to 20 epochs with early stopping. Gene names were converted to gene symbols where necessary. Spatial neighbor graphs were constructed using Delaunay triangulation with dataset-specific radius filtering. We treated the graph-level representations as niche embeddings.

**PASTE**. We ran PASTE in a pairwise manner using the center_align function with 15 NMF components and alpha=0.1. For each sample pair, NMF was applied to the center slice to obtain the W matrix as the embedding for the first sample. For the second sample, we projected its expression matrix onto the NMF basis via the pseudo-inverse of the H matrix. We treated the resulting NMF representations as niche embeddings.

**scNiche**. For the simulation study, we trained scNiche on all samples in a single run using process_multi_slices to ensure KNN neighbors were computed within each sample separately. Both cell neighborhood composition and expression-based views were used as input.

**STalign**. We ran STalign in a pairwise manner. For each sample pair, we rasterized expression-weighted density maps and applied the LDDMM (Large Deformation Diffeomorphic Metric Mapping) algorithm to learn a diffeomorphic spatial transformation from the source sample to the target sample. We then transformed all query spot coordinates using the learned transformation. For each query niche, we computed the matching score for each reference spot as a Gaussian kernel of the Euclidean distance to the nearest transformed niche spot.

**STAligner**. We trained STAligner on all samples in a single run. Highly variable genes were selected per sample, and their union was used as the common feature set. Sample-specific spatial networks were constructed with dataset-appropriate radius cutoffs. We treated the latent node embeddings from the graph attention network as niche embeddings.

For all methods, cosine similarity between embeddings was used to compute the niche matching score. All experiments were conducted on an NVIDIA A100 GPU with a memory size of 80 GB.

## Evaluation metrics

### Region matching score

To provide a ground-truth proxy for niche similarity based on cell type composition and topological structure, we adopted a graph-based approach based on the Wasserstein Weisfeiler-Lehman (WWL) graph kernel[76]. This metric, referred to as the region matching (RM) score, was used to evaluate how well the predicted niche similarities reflected known categorical cell type patterns.

Consider a global graph $G = (V, E, A)$ with categorical annotations $A$ as the node features. Given two niches modeled as two subgraphs $g_1 = (V_1, E_1, A_1)$ and $g_2 = (V_2, E_2, A_2)$, WWL performs message passing on these subgraphs in the following way:

$$a^{h+1}(v) = \text{hash}\left(a^h(v), \mathcal{N}^h(v)\right) \quad (14)$$

where $a^h(v)$ is the integer node label of $v$ after the $h$-th iteration, with $a^0(v)$ being the integer label converted from the original categorical cell type label. $\mathcal{N}^h(v)$ denotes the neighbor node labels of $v$ after the $h$-th iteration. The `hash` function takes the node labels of the last iteration involved in node $v$'s neighborhood as input and outputs a new integer node label as the updated node label. Specifically, we first constructed a string representation of the input node labels:

$$s^h(v) = \text{sort}\left(\text{concatenate}\left(a^h(v), \mathcal{N}^h(v)\right)\right) \quad (15)$$

where the integer node labels are concatenated and sorted in ascending order. Meanwhile, we maintained a global string hash table for each iteration: $\mathcal{D}^h\colon \text{str} \rightarrow \mathbb{N}$. This hash table maps each unique string representation encountered during the iteration to a unique integer ID, assigned in the order of appearance. The ID corresponds to the total number of distinct string representations seen so far in that iteration. The updated node label of $v$ is then computed as

$$a^{h+1}(v) = \mathcal{D}^h\left(s^h(v)\right) \quad (16)$$

We set the number of iterations $K = 3$ in all our experiments. After the message passing is done, we extract the final representation vector for node $v$ by concatenating all intermediate node labels:

$$f(v) = \text{concatenate}\left(a^0(v), a^1(v), \dots, a^K(v)\right) \quad (17)$$

Then for the two niches $g_1$ and $g_2$, we compute the Wasserstein distance between their feature vectors:

$$W(g_1, g_2) = \min_{\gamma} \sum_{v \in V_1} \sum_{u \in V_2} \gamma_{ij} \cdot \text{cost}\big(f(v), f(u)\big) \tag{18}$$

where $\gamma$ is the transport plan between the uniform distributions on the node feature vectors of the two subgraphs, and the cost function is chosen as Hamming distance since the graphs are categorically labeled. Finally, we defined the Region Matching (RM) Score by converting the distance metric into a similarity measure:

$$\text{RM}(g_1, g_2) = 1 - W(g_1, g_2) \tag{19}$$

For a niche querying scenario, we have a query NOI $\text{Niche}_q$ and a reference niche set $\mathcal{R}$ which contained the candidate niches on all reference samples. Let $x_r = \text{RM}\big(\text{Niche}_q, \text{Niche}_r\big), y_r = \text{matching_score}_{\text{Method}}(\text{Niche}_q, \text{Niche}_r), \text{Niche}_r \in \mathcal{R}$. We used the following Pearson Correlation Coefficient (PCC) metric to measure how the predicted matching score of different method resembled the RM score:

$$\text{PCC} = \frac{\sum_{\text{Niche}_r \in \mathcal{R}} (x_r - \bar{x})(y_r - \bar{y})}{\sqrt{\sum_{\text{Niche}_r \in \mathcal{R}} (x_r - \bar{x})^2} \sqrt{\sum_{\text{Niche}_r \in \mathcal{R}} (y_r - \bar{y})^2}} \tag{20}$$

where $\bar{x} = \frac{1}{|\mathcal{R}|} \sum_{\text{Niche}_r \in \mathcal{R}} x_r$ , $\bar{y} = \frac{1}{|\mathcal{R}|} \sum_{\text{Niche}_r \in \mathcal{R}} y_r$.

Based on this score, we developed RM-Ideal, where the input categorical labels were ground truth annotations, and RM-GraphST, which took GraphST's unsupervised clustering labels as input. The implementation was based on GraphCompass[77].

### Niche composition Jensen-Shannon score

To measure the compositional similarity between niches, we model the composition of each niche as a multinomial distribution and compute the Jensen-Shannon distance between the distributions of different niches. Specifically, let $P = (p_1, p_2, \dots, p_m), Q = (q_1, q_2, \dots, q_m)$ be the normalized compositional vector of Niche A and Niche B, where

$$\sum_{i=1}^{n} p_i = 1, \quad \sum_{i=1}^{n} q_i = 1, \quad p_i \geq 0, \quad q_i \geq 0 \tag{21}$$

We treat the two compositional vectors as two probabilistic distributions, and compute the Jensen-Shannon distance between them:

$$D_{\text{JS}}(P|\text{Q}) = \frac{1}{2}\sum_{i=1}^{n} p_i \log_2\left(\frac{p_i}{m_i}\right) + \frac{1}{2}\sum_{i=1}^{n} q_i \log_2\left(\frac{q_i}{m_i}\right) \tag{22}$$

where $m_i = \frac{1}{2}(p_i + q_i)$. We then reverse the distance measure into a niche composition Jensen-Shannon (NCJS) score:

$$\text{NCJS}(\text{Niche}_A, \text{Niche}_B) = 1 - D_{\text{JS}}(P|\text{Q}) \tag{23}$$

We also define the Average NCJS score to assess whether niches with similar compositions are embedded close to each other in the latent space of different methods. Given a dataset with $N$ niches, for each individual niche $i$, we first obtain its $k$ −nearest neighbors according to the cosine similarity between niche embeddings, denoting as $KNN(i)$ as the $k$ nearest neighbors of the $i$-th niche. We compute the following Neighborhood NSJS score for each individual niche:

$$\text{Neighborhood_NCJS}(i) = \frac{1}{k}\sum_{j \in \text{KNN}(i)} \text{NCJS}\left(\text{Niche}_i, \text{Niche}_j\right) \tag{24}$$

We then aggregate the Neighborhood NSJS over all niches as the Average NCJS score:

$$\text{Average_NCJS} = \frac{1}{N}\sum_{i=1}^{N} \text{Neighborhood_NCJS}(i) \tag{25}$$

We set $k = 10$ in all our experiments.

## Downstream analysis

### Differential gene expression analysis

We calculated the differentially expressed genes (DEGs) using the "`rank_genes_groups`" function in Scanpy[75] Python package (version 1.9.8) with the Wilcoxon rank sum test. We set the threshold as adjusted *p*-value < 0.05 and log2FoldChange > 1.

### Cell type deconvolution analysis

We used cell2location[49] to perform cell type deconvolution on the ccRCC dataset with a reference scRNA-seq dataset[48]. We set `N_cells_per_location` as 30 and `detection_alpha` as 20. We obtained the q05 posterior quantiles as the estimated cell type proportions for each spot, and the number of samples from the posterior distributions was set as 1000.

### Survival analysis

The correlation between identified DEGs and survival outcomes in the JAVELIN 101 cohort was analyzed using the R packages survival[78] (version 3.8.3), survminer[79] (version 0.5.0), and survRM2[80] (version 1.0.4). Clinical data and normalized transcriptomics profiles of the JAVELIN 101 cohort were obtained from the original study[47]. Only patients treated with the combination of Avelumab and Axitinib were included in the analysis. Individual patient enrichment scores of each DEG set (consisting of the 10 DEGs with the highest log2FoldChange) were determined using single-sample gene set enrichment analysis (ssGSEA) via the R package GSVA[81] (v.2.0.7). Patients with high ssGSEA scores for the TLS-related DEG set were subsequently selected for further survival analysis using the DEG sets associated with Types A and B. Similarly, patients with high ssGSEA scores for the Type B-related DEG set were further selected to perform survival analysis with Type B1 and B2 DEG sets. Group-wise survival differences were assessed using restricted mean survival time (RMST) analysis at a truncation time of $\tau = 15$ months, computed with the survRM2::rmst2 function.

### Gene set enrichment analysis

We performed the gene set enrichment analysis[82] (GSEA) using the `enrichr` function of the gseapy[83] Python package (version 1.1.3) with default parameters. We analyzed the upregulated gene list against the MSigDB Hallmark 2020[84] and KEGG_2019_Human[85] database.

## Spatial organization analysis

To examine cellular composition and pathway activity as a function of TLS depth, we restricted this analysis to a curated set of ccRCC TLS. A TLS is qualified if (i) its total spot count was between 50 and 100, (ii) onion-peel decomposition yielded at least three layers, and (iii) the spots at layer 3 or deeper, summed across all such layers, were at least 3 in number. Layer assignment was performed as follows: the outer layer L1 was defined as the TLS spots whose set of graph neighbors was not fully contained within the TLS; those spots were removed, and the procedure was repeated on the remaining set to define L2, L3, and so on. These layers were then collapsed into three positional bins: Outer (L1), Middle (L2), and Inner (L3 or deeper merged).

For each cell-type proportion and pathway score, the displayed line plots show, for each TLS group (A, B1, B2) at each position (Inner, Middle, Outer), the mean value across all spots in that positional bin. Cell-type proportions per spot were taken from the deconvolution results from cell2location. Pathway scores per spot were computed with Scanpy's `score_genes` function on library-size-normalized (target sum 10,000), log1p-transformed counts. The score is the mean expression of the target gene set minus the mean expression of a control gene set randomly drawn from the same expression bins. Pathway gene sets were retrieved from MSigDB Hallmark 2020 and KEGG_2019_Human via gseapy.

## Cross-cancer-type analysis

Building on the QueST model pretrained on the NSCLC CosMx dataset, we transferred the three TLS subtypes discovered in ccRCC (Type A, B1, B2) onto NSCLC and analyzed the fibroblastic identity of the resulting cross-cancer niches. For each ccRCC slice, we restricted the gene panel to the 895 genes used by the NSCLC-pretrained encoder and obtained a 32-dimensional embedding for each TLS by mean-pooling the encoder's node embeddings within the TLS mask. For each NSCLC candidate niche and each ccRCC TLS subtype (A, B1, or B2), we computed a subtype matching score as the mean cosine similarity between the niche embedding and the embeddings of all ccRCC TLS belonging to that subtype, yielding three matching scores per spot. We then took the maximum of its three subtype matching scores. Within each NSCLC slice, the niches whose maximum subtype score fell in the top 10% were retained as the queried TLS candidate region. Within this candidate region, each niche was further assigned to the subtype with the highest matching score, yielding three types of queried TLS regions (Type-A, Type-B1, and Type-B2).

To examine whether the fibrotic signature that characterizes Type-A TLS in ccRCC is conserved at the single-cell level in NSCLC, we derived a cross-cancer consensus CAF gene signature from the intersection of two independent differential expression analyses. On ccRCC, we computed DEGs between Type-A and Type-B TLS and obtained upregulated genes of Type-A TLS. On NSCLC, we extracted cells annotated as fibroblasts and computed DEGs between fibroblasts located within Type-A regions and all remaining fibroblasts. The intersection of the two upregulated gene sets, restricted to the threshold at log2 fold-change > 0.5 and adjusted $p < 0.01$ on both sides, yielded 38 cross-cancer upregulated genes. These were further filtered through a curated CAF gene-category dictionary compiled from canonical CAF reviews and CAF subtyping studies, retaining 22 genes spanning six functional categories as the cross-cancer consensus CAF signature. We then computed the CAF score using the "score_genes" function in Scanpy on normalized expression. For each cell, the score is the mean expression of the target gene set minus the mean expression of a control gene set randomly drawn from the same expression bins.

### *De novo* NOI identification

Denote $G$ as the global niche set, containing niches of all samples including the source sample, and $S$ as the source sample niche set. We define niche patterns as niche cluster subsets of $G$. For an arbitrary niche $n \in S$, let $P(n)$ be the niche pattern which $n$ belongs to. We then define a niche prevalence ratio (NPR) as follows:

$$\text{NPR}(n) = \frac{|P(n)|/|G|}{|P(n) \cap S|/|S|}$$

where the $|\cdot|$ operator represents the number of niches contained in the corresponding niche set. Intuitively, the ratio of these two proportions yields a score much greater than 1 when the corresponding niche pattern is common across the dataset but constitutes only a small minority in the source sample.

### Nodule classification

We reproduced the tumor nodule phenotype classification from the original study to enable downstream analysis of queried nodules. The classification assigns each nodule to one of five tumor-intrinsic phenotypes: portal, central, cholangiocytic, Hamp2/Upp2-enriched, and histone-enriched, and Unclassified if no phenotype passes the classification threshold. The procedure is as follows: for each nodule, we computed the mean expression of phenotype-specific marker genes across all spots belonging to that nodule. The marker genes for each phenotype were: portal (Sds, Sdsl, Hal), central (Oat, Gulo, Cyp2e1, Cyp1a2), cholangiocytic (Krt19, Krt7, Epcam, Gp2), Hamp2/Upp2-enriched (Hamp, Hamp2, Upp2, Car3), and histone-enriched (Hist1h1b, Hist1h2ae, Hist1h2ap, Hist1h3b). Expression values were computed after library-size normalization (target sum = 10,000) and log1p transformation. To enable cross-gene comparison, mean expression values were normalized using the 25th and 99th percentiles (q25, q99) computed across all nodules from 11 tissue sections (6 main sections and 5 technical replicates). For each phenotype, a nodule was assigned a positive binary label (bin = 1) if at least 2 of its marker genes had normalized values exceeding 0.5.

## References

1. Zhang, B. & Chen, T. Local and systemic mechanisms that control the hair follicle stem cell niche. *Nat Rev Mol Cell Biol* **25**, 87–100 (2024).

2. Kanemaru, K. *et al.* Spatially resolved multiomics of human cardiac niches. *Nature* **619**, 801–810 (2023).

3. Chen, J., Larsson, L., Swarbrick, A. & Lundeberg, J. Spatial landscapes of cancers: insights and opportunities. *Nat Rev Clin Oncol* **21**, 660–674 (2024).

4. Jain, S. & Eadon, M. T. Spatial transcriptomics in health and disease. *Nat Rev Nephrol* **20**, 659–671 (2024).

5. Xia, C., Fan, J., Emanuel, G., Hao, J. & Zhuang, X. Spatial transcriptome profiling by MERFISH reveals subcellular RNA compartmentalization and cell cycle-dependent gene expression. *Proceedings of the National Academy of Sciences* **116**, 19490–19499 (2019).

6. Eng, C.-H. L. *et al.* Transcriptome-scale super-resolved imaging in tissues by RNA seqFISH+. *Nature* **568**, 235–239 (2019).

7. Rao, N., Clark, S. & Habern, O. Bridging Genomics and Tissue Pathology. *Genetic Engineering & Biotechnology News* **40**, 50–51 (2020).

8. Rodriques, S. G. *et al.* Slide-seq: A scalable technology for measuring genome-wide expression at high spatial resolution. *Science* **363**, 1463–1467 (2019).

9. Chen, A. *et al.* Spatiotemporal transcriptomic atlas of mouse organogenesis using DNA nanoball-patterned arrays. *Cell* **185**, 1777-1792.e21 (2022).

10. Arora, R. *et al.* Spatial transcriptomics reveals distinct and conserved tumor core and edge architectures that predict survival and targeted therapy response. *Nat Commun* **14**, 5029 (2023).

11. Schürch, C. M. *et al.* Coordinated Cellular Neighborhoods Orchestrate Antitumoral Immunity at the Colorectal Cancer Invasive Front. *Cell* **182**, 1341-1359.e19 (2020).

12. Baccin, C. *et al.* Combined single-cell and spatial transcriptomics reveal the molecular, cellular and spatial bone marrow niche organization. *Nat Cell Biol* **22**, 38–48 (2020).

13. Zhao, J. J. *et al.* Spatially Resolved Niche and Tumor Microenvironmental Alterations in Gastric Cancer Peritoneal Metastases. *Gastroenterology* **167**, 1384-1398.e4 (2024).

14. Bagaev, A. *et al.* Conserved pan-cancer microenvironment subtypes predict response to immunotherapy. *Cancer Cell* **39**, 845-865.e7 (2021).

15. Shi, Q. *et al.* Cross-tissue multicellular coordination and its rewiring in cancer. *Nature* https://doi.org/10.1038/s41586-025-09053-4 (2025) doi:10.1038/s41586-025-09053-4.

16. Birk, S. *et al.* Quantitative characterization of cell niches in spatially resolved omics data. *Nat Genet* https://doi.org/10.1038/s41588-025-02120-6 (2025) doi:10.1038/s41588-025-02120-6.

17. Qian, J. *et al.* Identification and characterization of cell niches in tissue from spatial omics data at single-cell resolution. *Nat Commun* **16**, 1693 (2025).

18. Dong, K. & Zhang, S. Deciphering spatial domains from spatially resolved transcriptomics with an adaptive graph attention auto-encoder. *Nat Commun* **13**, 1739 (2022).

19. Ren, H., Walker, B. L., Cang, Z. & Nie, Q. Identifying multicellular spatiotemporal organization of cells with SpaceFlow. *Nat Commun* **13**, 4076 (2022).

20. Guo, T. *et al.* SPIRAL: integrating and aligning spatially resolved transcriptomics data across different experiments, conditions, and technologies. *Genome Biol* **24**, 241 (2023).

21. Long, Y. *et al.* Spatially informed clustering, integration, and deconvolution of spatial transcriptomics with GraphST. *Nat Commun* **14**, 1155 (2023).

22. Tang, Z. *et al.* Search and match across spatial omics samples at single-cell resolution. *Nat Methods* https://doi.org/10.1038/s41592-024-02410-7 (2024) doi:10.1038/s41592-024-02410-7.

23. Zhou, X., Dong, K. & Zhang, S. Integrating spatial transcriptomics data across different conditions, technologies and developmental stages. *Nat Comput Sci* **3**, 894–906 (2023).

24. Xia, C.-R., Cao, Z.-J., Tu, X.-M. & Gao, G. Spatial-linked alignment tool (SLAT) for aligning heterogenous slices. *Nat Commun* **14**, 7236 (2023).

25. Clifton, K. *et al.* STalign: Alignment of spatial transcriptomics data using diffeomorphic metric mapping. *Nat Commun* **14**, 8123 (2023).

26. Zeira, R., Land, M., Strzalkowski, A. & Raphael, B. J. Alignment and integration of spatial transcriptomics data. *Nat Methods* **19**, 567–575 (2022).

27. Levy, N., Nadler, B., Ingelfinger, F. & Bakulin, A. NICHEVI: A PROBABILISTIC FRAMEWORK TO EM- BED CELLULAR INTERACTION IN SPATIAL TRANSCRIPTOMICS. (2024).

28. Varrone, M., Tavernari, D., Santamaria-Martínez, A., Walsh, L. A. & Ciriello, G. CellCharter reveals spatial cell niches associated with tissue remodeling and cell plasticity. *Nat Genet* https://doi.org/10.1038/s41588-023-01588-4 (2023) doi:10.1038/s41588-023-01588-4.

29. Hu, Y. *et al.* Unsupervised and supervised discovery of tissue cellular neighborhoods from cell phenotypes. *Nat Methods* https://doi.org/10.1038/s41592-023-02124-2 (2024) doi:10.1038/s41592-023-02124-2.

30. Wu, Z. *et al.* Graph deep learning for the characterization of tumour microenvironments from spatial protein profiles in tissue specimens. *Nat. Biomed. Eng* **6**, 1435–1448 (2022).

31. Fischer, D. S., Ali, M., Richter, S., Ertürk, A. & Theis, F. *Graph Neural Networks Learn Emergent Tissue Properties from Spatial Molecular Profiles*. http://biorxiv.org/lookup/doi/10.1101/2022.12.08.519537 (2022) doi:10.1101/2022.12.08.519537.

32. Wang, H. *et al.* Neural Similarity Search on Supergraph Containment. *IEEE Trans. Knowl. Data Eng.* **36**, 281–295 (2024).

33. Wu, R. *et al.* Comprehensive analysis of spatial architecture in primary liver cancer. *SCIENCE ADVANCES* (2021).

34. Korsunsky, I. *et al.* Fast, sensitive and accurate integration of single-cell data with Harmony. *Nat Methods* **16**, 1289–1296 (2019).

35. Goodfellow, I. *et al.* Generative adversarial networks. *Commun. ACM* **63**, 139–144 (2020).

36. Qian, J. *et al.* Simulating multiple variability in spatially resolved transcriptomics with scCube. *Nat Commun* **15**, 5021 (2024).

37. Maynard, K. R. *et al.* Transcriptome-scale spatial gene expression in the human dorsolateral prefrontal cortex. *Nat Neurosci* **24**, 425–436 (2021).

38. Luecken, M. D. *et al.* Benchmarking atlas-level data integration in single-cell genomics. *Nat Methods* **19**, 41–50 (2022).

39. Teillaud, J.-L., Houel, A., Panouillot, M., Riffard, C. & Dieu-Nosjean, M.-C. Tertiary lymphoid structures in anticancer immunity. *Nat Rev Cancer* **24**, 629–646 (2024).

40. Sautès-Fridman, C., Petitprez, F., Calderaro, J. & Fridman, W. H. Tertiary lymphoid structures in the era of cancer immunotherapy. *Nat Rev Cancer* **19**, 307–325 (2019).

41. Schumacher, T. N. & Thommen, D. S. Tertiary lymphoid structures in cancer. *Science* **375**, eabf9419 (2022).

42. Xu, W. *et al.* Heterogeneity in tertiary lymphoid structures predicts distinct prognosis and immune microenvironment characterizations of clear cell renal cell carcinoma. *J Immunother Cancer* **11**, e006667 (2023).

43. Zhang, Y. *et al.* Tertiary lymphoid structural heterogeneity determines tumour immunity and

prospects for clinical application. *Mol Cancer* **23**, 75 (2024).

44. Yan, Y. *et al.* Multi-omic profiling highlights factors associated with resistance to immuno-chemotherapy in non-small-cell lung cancer. *Nat Genet* **57**, 126–139 (2025).

45. Tang, Z. *et al.* Spatial transcriptomics reveals tryptophan metabolism restricting maturation of intratumoral tertiary lymphoid structures. *Cancer Cell* S1535610825001126 (2025) doi:10.1016/j.ccell.2025.03.011.

46. Meylan, M. *et al.* Tertiary lymphoid structures generate and propagate anti-tumor antibody-producing plasma cells in renal cell cancer. *Immunity* **55**, 527-541.e5 (2022).

47. Motzer, R. J. *et al.* Avelumab plus axitinib versus sunitinib in advanced renal cell carcinoma: biomarker analysis of the phase 3 JAVELIN Renal 101 trial. *Nat Med* **26**, 1733–1741 (2020).

48. Li, R. *et al.* Mapping single-cell transcriptomes in the intra-tumoral and associated territories of kidney cancer. *Cancer Cell* **40**, 1583-1599.e10 (2022).

49. Kleshchevnikov, V. *et al.* Cell2location maps fine-grained cell types in spatial transcriptomics. *Nat Biotechnol* **40**, 661–671 (2022).

50. Salmon, H. *et al.* Matrix architecture defines the preferential localization and migration of T cells into the stroma of human lung tumors. *J. Clin. Invest.* **122**, 899–910 (2012).

51. Joyce, J. A. & Fearon, D. T. T cell exclusion, immune privilege, and the tumor microenvironment. *Science* **348**, 74–80 (2015).

52. Mariathasan, S. *et al.* TGFβ attenuates tumour response to PD-L1 blockade by contributing to exclusion of T cells. *Nature* **554**, 544–548 (2018).

53. Chen, D. S. & Mellman, I. Elements of cancer immunity and the cancer–immune set point. *Nature* **541**, 321–330 (2017).

54. Li, X. *et al.* Integrated spatial transcriptomic profiling to dissect the cellular characteristics of tumor-associated tertiary lymphoid structures. *Cell Reports* **44**, 116250 (2025).

55. Lodi, F. *et al.* Decoding tumor heterogeneity: A spatially informed pan-cancer analysis of the tumor microenvironment. *Cell Reports Medicine* **6**, 102416 (2025).

56. Vanhersecke, L. *et al.* Mature tertiary lymphoid structures predict immune checkpoint inhibitor efficacy in solid tumors independently of PD-L1 expression. *Nat Cancer* **2**, 794–802 (2021).

57. Liu, Y. *et al.* Conserved spatial subtypes and cellular neighborhoods of cancer-associated fibroblasts revealed by single-cell spatial multi-omics. *Cancer Cell* S1535610825000832 (2025) doi:10.1016/j.ccell.2025.03.004.

58. Sun, J. *et al.* Exploring the cross-cancer effect of circulating proteins and discovering potential intervention targets for 13 site-specific cancers. *JNCI: Journal of the National Cancer Institute* **116**, 565–573 (2024).

59. He, S. *et al.* High-plex imaging of RNA and proteins at subcellular resolution in fixed tissue by spatial molecular imaging. *Nat Biotechnol* **40**, 1794–1806 (2022).

60. Breinig, M. *et al.* Integrated in vivo combinatorial functional genomics and spatial transcriptomics of tumours to decode genotype-to-phenotype relationships. *Nat. Biomed. Eng* **10**, 125–143 (2025).

61. Bian, H., Chen, Y., Wei, L. & Zhang, X. uHAF: a unified hierarchical annotation framework for cell type standardization and harmonization. *Bioinformatics* **41**, btaf149 (2025).

62. Börner, K. *et al.* Human BioMolecular Atlas Program (HuBMAP): 3D Human Reference Atlas Construction and Usage. Preprint at https://doi.org/10.1101/2024.03.27.587041 (2024).

63. Walker, B. L. & Nie, Q. NeST: nested hierarchical structure identification in spatial transcriptomic data. *Nat Commun* **14**, 6554 (2023).

64. Hao, M. *et al.* Large-scale foundation model on single-cell transcriptomics. *Nat Methods* **21**, 1481–1491 (2024).

65. Bian, H. *et al.* scMulan: A Multitask Generative Pre-Trained Language Model for Single-Cell Analysis. in *Research in Computational Molecular Biology* (ed. Ma, J.) vol. 14758 479–482 (Springer Nature Switzerland, Cham, 2024).

66. Yang, F. *et al.* scBERT as a large-scale pretrained deep language model for cell type annotation of single-cell RNA-seq data. *Nat Mach Intell* **4**, 852–866 (2022).

67. Cui, H. *et al.* scGPT: toward building a foundation model for single-cell multi-omics using generative AI. *Nat Methods* **21**, 1470–1480 (2024).

68. Theodoris, C. V. *et al.* Transfer learning enables predictions in network biology. *Nature* **618**, 616–624 (2023).

69. Wen, H. *et al. CellPLM: Pre-Training of Cell Language Model Beyond Single Cells*. http://biorxiv.org/lookup/doi/10.1101/2023.10.03.560734 (2023) doi:10.1101/2023.10.03.560734.

70. Wang, C. *et al.* scGPT-spatial: Continual Pretraining of Single-Cell Foundation Model for Spatial Transcriptomics.

71. Hao, M. *et al.* GeST: Towards Building A Generative Pretrained Transformer for Learning Cellular Spatial Context.

72. Palla, G. *et al.* Squidpy: a scalable framework for spatial omics analysis. *Nat Methods* **19**, 171–178 (2022).

73. Xu, K., Hu, W., Leskovec, J. & Jegelka, S. How Powerful are Graph Neural Networks? in *International Conference on Learning Representations* (2019).

74. Veličković, P. *et al.* Deep Graph Infomax. Preprint at http://arxiv.org/abs/1809.10341 (2018).

75. Wolf, F. A., Angerer, P. & Theis, F. J. SCANPY: large-scale single-cell gene expression data analysis. *Genome Biol* **19**, 15 (2018).

76. Togninalli, M., Ghisu, E., Llinares-López, F., Rieck, B. & Borgwardt, K. Wasserstein Weisfeiler-Lehman Graph Kernels.

77. Ali, M. *et al. GraphCompass: Spatial Metrics for Differential Analyses of Cell Organization across Conditions*. http://biorxiv.org/lookup/doi/10.1101/2024.02.02.578605 (2024) doi:10.1101/2024.02.02.578605.

78. Therneau, T. A package for survival analysis in R.

79. Alboukadel, K., Marcin, K. & Przemyslaw, B. survminer: Drawing Survival Curves using ‘ggplot2’. (2024).

80. Hajime Uno, Lu Tian, Miki Horiguchi, Angel Cronin, Chakib Battioui, James Bell. survRM2: Comparing Restricted Mean Survival Time. 1.0-4 https://doi.org/10.32614/CRAN.package.survRM2 (2015).

81. Hänzelmann, S., Castelo, R. & Guinney, J. GSVA: gene set variation analysis for microarray and RNA-Seq data. *BMC Bioinformatics* **14**, 7 (2013).

82. Subramanian, A. *et al.* Gene set enrichment analysis: A knowledge-based approach for interpreting genome-wide expression profiles. *Proc. Natl. Acad. Sci. U.S.A.* **102**, 15545–15550 (2005).

83. Fang, Z., Liu, X. & Peltz, G. GSEApy: a comprehensive package for performing gene set enrichment analysis in Python. *Bioinformatics* **39**, btac757 (2023).

84. Liberzon, A. *et al.* The Molecular Signatures Database Hallmark Gene Set Collection. *Cell Systems* **1**, 417–425 (2015).

85. Kanehisa, M., Sato, Y., Furumichi, M., Morishima, K. & Tanabe, M. New approach for understanding genome variations in KEGG. *Nucleic Acids Research* **47**, D590–D595 (2019).

## Data and Code availability

The codes of the QueST model and codes for reproducing the results in this paper are available at https://github.com/cmhimself/QueST. The tutorials for using the QueST model are available at https://quest-niche.readthedocs.io/en/latest/index.html. The DLPFC dataset was downloaded from http://spatial.libd.org/spatialLIBD/. The Mouse Olfactory Bulb Tissue dataset was downloaded from https://github.com/guott15/SPIRAL_pyG. The ccRCC dataset was obtained from GEO under accession GSE175540. The NSCLC dataset was downloaded from https://github.com/Lotfollahi-lab/nichecompass-reproducibility. The Perturb-CAST dataset was downloaded from https://zenodo.org/records/10986436.

# Supplementary Materials

Supplementary Note 1-5, Fig. S1-12, Table S1.

## Supplementary Notes

### Supplementary Note 1: distinction between the niche-query task and previous niche analytical paradigms

Previous methods, including NicheCompass, mainly worked on unsupervised niche identification and reference mapping, which is closely related with the niche-query task. However, these tasks have fundamental differences in both the problem setting and the application scenarios. Here, we provide a detailed clarification on these essential distinctions.

NicheCompass did a great job in quantitatively characterizing cell niches in spatial omics data. They took a "cell-centric" approach to learn interpretable cell embeddings that encode signaling events in the close neighborhood of an individual cell, which is defined as "a cell's niche" which represent the local microenvironment shaped by the signals it receives and sends. They distinguished "neighborhood component" vs. "self component" in defining the niche. This allows the method to embed every cell in the tissue according to its niche characteristics and then cluster these embeddings to identify larger tissue segments composed of different types of niches. In this way, they successfully resolved the global mapping of tissue architecture in terms of cell-cell communication characteristics, revealing the diversity and hierarchy of niches across the tissue. In fact, the term "niche" has two meanings in that work: One is "a cell's niche" which captures the cell's communications with immediate neighboring cells, and the other is the larger regions in the tissue identified by the clustering of cells' niches that share the same functional characteristics.

The "niche" we studied in the niche-query task has a very different definition. It refers to a small conserved local area in the tissue that has specific patterns in its cell composition and topological structure. Such special patterns can have conserved functions across different tissues or disease contexts, such as the TLS in different types of cancers. We took a "niche-centric" view to model such niches in the sense that we do not define it as the neighborhood around any individual cell, but define it as the structure and functional composition of multiple equally-important cells. A niche is the unit of analysis itself, not anchored on a specific cell. The niches with structural or functional importance are referred to as niches of interest or NOI. The niche-query task is to search for the occurrence of NOIs in target samples when users have particular NOIs they want to study. Methods working well on this task can also be used to search across multiple samples to identify possible *de novo* NOIs shared in the samples, which could help to discover previously unknown functional niches. Such niches usually do not cover an entire region of the tissue, but are sparse local functional structures that play key physiological or pathological roles. While the niche-related tasks existing methods work on have their own significance, this new niche-query task has its own biological significance in identifying local functional sub-structures, especially for disease studies.

**Supplementary Note 2: design rationale and methodological contribution of QueST**

As we discussed in the Introduction section, the nature of the niche-query task requires three key properties (central-cell invariance, strict locality, and structure awareness) that methods for this task should possess. Existing embedding-based niche-analysis methods were not designed for the niche-query task. They have the flexibility to be adopted for the new task, but they are not explicitly designed to enforce these key properties. In particular, cell-centric representations approximate local context through message passing and do not take the niche itself as the object, making them difficult to enforce locality at the niche boundary or to directly encode internal spatial patterns. These are the main reasons behind their observed inferior performance in our experiments reported in the previous manuscript as well as those newly conducted in this revision.

QueST is designed specifically for the niche-query task. QueST takes a niche-centric representation learning strategy by explicitly modeling each niche as a subgraph and learns representations at the level of these structured units. The model architecture and training objectives directly align with the requirements of the niche-query task: subgraph-based modeling enforces locality and central-cell invariance, and our newly designed contrastive objectives promote discrimination between niches with similar composition but different spatial organization, enabling structure awareness. While some components used in the model (graph autoencoders, adversarial training) are individually well-established and are not our inventions, the proper integration of these existing modules yielded superior performance on both simulation data and real application data. These results show the importance of the alignment between task setting and method design.

### Supplementary Note 3: distinction between niche querying and spatial alignment

Spatial alignment methods can be broadly categorized into two paradigms based on their assumptions about cross-sample similarity:

**Homogeneous alignment methods** (e.g., PASTE, STAligner, STalign) assume that the two samples to be aligned share global similarity in tissue architectures. Under this assumption, they seek a global mapping between samples, typically by optimizing a spatial correspondence that preserves overall tissue organization. This design is suited for aligning replicate sections or serial slices of the same tissue, but becomes problematic when the source and target samples have different tissue backgrounds as in the setting of niche querying across samples. For example, in the simulation experiments where two samples share local niche types but differ in background tissue composition (endothelial vs. macrophage/monocyte), homogeneous alignment methods lack a meaningful global correspondence to exploit.

**Heterogeneous alignment methods** (e.g., SLAT, CAST) relax the global similarity assumption and are capable of aligning samples with different tissue compositions by establishing cell-to-cell correspondences based on local features. While this represents a step toward handling cross-sample heterogeneity, these methods still operate at the cell level: they are optimized for establishing correspondences rather than selectively retrieving spatially organized regions. This introduces a limitation for niche-level tasks: they do not explicitly model niches as integrated spatial units, and therefore struggle to distinguish regions that share similar cell type composition but differ in spatial organization.

We illustrated this with the T-edge query in the simulation experiments. CAST misaligned query NOIs to regions with mismatched cell types due to its reliance on global alignment (Fig. S2a). SLAT aligned the T cells and cancer/epithelial cells in the query NOI to corresponding cell types in the target sample. However, because SLAT operates at the cell level without modeling the collective spatial arrangement of these cells, it indiscriminately aligned the query NOI to all niches containing both T cells and cancer cells, including T-core and B-edge niches (Fig. S2b). In contrast, QueST models each niche as an integrated subgraph whose embedding encodes spatial topology, enabling it to correctly retrieve only T-edge niches while rejecting topologically dissimilar alternatives.

In summary, the niche-query task differs from spatial alignment in both scope and granularity. Homogeneous alignment methods are restricted to globally similar samples and thus cannot operate across diverse tissue backgrounds. Heterogeneous alignment methods can handle background diversity but remain limited to cell-level correspondences, which are insufficient for capturing niche-level topological distinctions. QueST addresses both limitations by directly modeling niches as a cohesive unit and learning niche-level representations that encode compositional and topological information jointly.

**Supplementary Note 4: sensitivity analysis and ablation study of QueST**

**Sensitivity analyses.** We assessed the sensitivity of QueST to two key design choices: the subgraph hop number $K$ and the GNN backbone architecture, both evaluated on the DLPFC dataset.

For the subgraph hop number $K$, we varied it from 1 to 5 while keeping all other settings fixed. As shown in Fig. S6a, QueST maintains stable performance across all values of $K$ and all 12 NOI types, indicating low sensitivity to this hyperparameter. The default choice of $K = 3$ is motivated by prior work on subgraph-based modeling of cellular microenvironments. For example, the SPACE-GM framework adopts 3-hop local subgraphs to define microenvironments for supervised spatial proteomic data classification, and shows that this scale provides a good balance between capturing sufficient local context and maintaining computational efficiency. We followed this convention in QueST. For the GNN backbone, we compared Graph Isomorphism Network (GIN; default in QueST), Graph Convolutional Neural Network (GCN), and Graph Attention Network (GAT) (Fig. S6b).

QueST achieves consistently robust performance across all tested values of $K$ and all GNN architectures, suggesting that its advantage is primarily attributable to its overall framework design: subgraph modeling, contrastive learning, and adversarial training.

**Ablation study.** We evaluated six ablated variants on the DLPFC dataset by removing or replacing individual components (Fig. S6c): (1) wo-Contrast, removing the contrastive learning module; (2) wo-Pooling, replacing the mean pooling step with the center node's embedding as the niche representation (as in cell-centric methods); (3) wo-Residual, removing residual connections between GNN layers in the encoder; (4) Euclidean, replacing cosine similarity with reversed Euclidean distance for computing niche matching scores; (5) wo-BatchAdv, removing the adversarial batch removal module; and (6) wo-Recon, removing the reconstruction loss.

The results reveal a clear hierarchy of component importance. The contrastive learning module is the most critical component, and its removal leads to the largest performance drop (median PCC from ~0.67 to ~0.21). This reflects its role in explicitly structuring the embedding space by separating distinct niches through negative pairs while enforcing locality through positive pairs. These are capabilities that reconstruction-based learning alone cannot provide. The subgraph-level mean pooling strategy is the second most important component. Replacing it with a center-node representation reduces performance substantially (median PCC from ~0.67 to ~0.46), highlighting the importance of treating niches as collective entities rather than single-cell proxies. Residual connections also contribute notably (median PCC ~0.51 without them), consistent with their role in mitigating over-smoothing in deep GNNs and preserving informative node representations within the encoder.

The remaining components, including cosine similarity (vs. Euclidean distance), adversarial batch removal, and reconstruction loss, provide moderate but consistent improvements. Cosine similarity measures the angular difference between embeddings regardless of their magnitude, making it more suitable for comparing niche representations that may vary in scale across samples. The adversarial batch removal module contributes to cross-sample robustness, and the reconstruction loss helps regularize the learned embeddings by encouraging them to retain biologically meaningful gene expression information.

Collectively, these results show that contrastive learning and subgraph-level pooling are the two dominant contributors, directly supporting QueST's core design principles of central-cell invariance, strict locality, and structure awareness, while the remaining components provide complementary gains.

**Supplementary Note 5: detailed explanation of cross-resolution niche querying**

QueST does not perform explicit resolution harmonization. Instead, it defines each candidate niche as a $K$-hop subgraph ($K = 3$) on the spatial graph and computes embeddings in a uniform manner across all datasets. Given a query NOI, QueST retrieves similar niches by computing cosine similarity between the NOI embedding and the embeddings of all candidate subgraphs in the target sample. This procedure is identical regardless of whether the source and target datasets share the same spatial resolution.

At the level of local niches, our assumption is not geometric equivalence, but that spatial structures with similar biological organization give rise to similar compositional and topological patterns, even when measured at different resolutions. For example, at a tissue boundary region, such as the GCL–MCL interface in the Mouse Olfactory Bulb (MOB), a local neighborhood at different resolutions may span different physical extents, but still reflects a similar transition pattern in terms of cell-type composition and relative spatial arrangement. In this sense, QueST compares niches based on these pattern-level features, rather than exact spatial scale. This is empirically supported by our MOB experiments, where three samples profiled with Visium (50 μm), Stereo-seq (35 μm), and Slide-seq V2 (10 μm) were used for cross-platform querying. Despite a five-fold difference in spatial resolution between Visium and Slide-seq V2, QueST achieved PCC values above 0.8 on both target samples (Fig. 4b).

For more complex structures such as TLS, QueST further improves robustness by representing each TLS as an aggregation of multiple local subgraph embeddings. Specifically, a TLS region is decomposed into a collection of overlapping $K$-hop subgraphs, and its final representation is obtained by pooling over these local embeddings. As a result, a TLS is modeled not as a single fixed-scale unit, but as a composite of multiple sub-niches, each capturing different aspects of its internal organization.

This aggregation is particularly important for cross-resolution comparison. For instance, in the Type-A TLSs observed in ccRCC samples, we consistently observed a fibroblast-enriched boundary region surrounding an immune-cell-present core. These regions represent distinct local patterns and give rise to different local embeddings. Even when profiled at different spatial resolutions, subgraphs corresponding to similar regions (e.g., core-like or boundary-like areas) tend to produce similar embeddings, while embeddings from different regions remain distinguishable. The pooling operation integrates these heterogeneous local representations, allowing the final TLS embedding to capture both the presence and the relative contribution of these substructures. As a result, the global embedding reflects the overall organizational pattern of the TLS. Therefore, two TLSs can be meaningfully compared based on their internal pattern composition, even when the underlying spatial resolution differs, and different TLS types can be distinguished according to their characteristic structural organization.

The cross-cancer TLS experiments supported this point. In this setting, ccRCC samples were profiled with Visium (50 μm, spot-level resolution), whereas NSCLC samples were profiled with CosMx at single-cell resolution, representing a substantial resolution gap. Despite this, QueST consistently achieved AUROC scores above 0.9 across all NSCLC samples. In addition, we observed consistent structural and functional patterns across datasets. For example, we recapitulated the tumor proximity and fibroblast encapsulation pattern of Type-A TLS in the NSCLC data. Cancer-associated fibroblast programs and EMT/ECM signatures were also upregulated in corresponding regions. Type-B1 niches were positioned farther from the tumor boundary, which is also consistent with ccRCC Type-B1 patterns. These cross-dataset consistencies further support that the learned embeddings capture biologically meaningful organizational patterns that are preserved across resolutions and cancer types.

We acknowledge that the use of a fixed $K$ is a limitation of the current design. In cases involving extreme resolution disparity or niches with highly heterogeneous internal structures, a fixed-scale subgraph representation may not fully capture fine-grained spatial organization. Our current strategy represents higher-order structures (e.g., TLS) by aggregating local subgraph embeddings into a single "super-niche" representation. While this pooling-based approach improves robustness to resolution differences, it does not explicitly model the spatial relationships between sub-niches, such as their relative arrangement or higher-order topology within the niche. As a result, certain structural information may be lost during aggregation. A promising direction to address these limitations is to extend QueST toward multi-scale and hierarchical niche modeling. We leave these work for future studies.

## Supplementary Figures

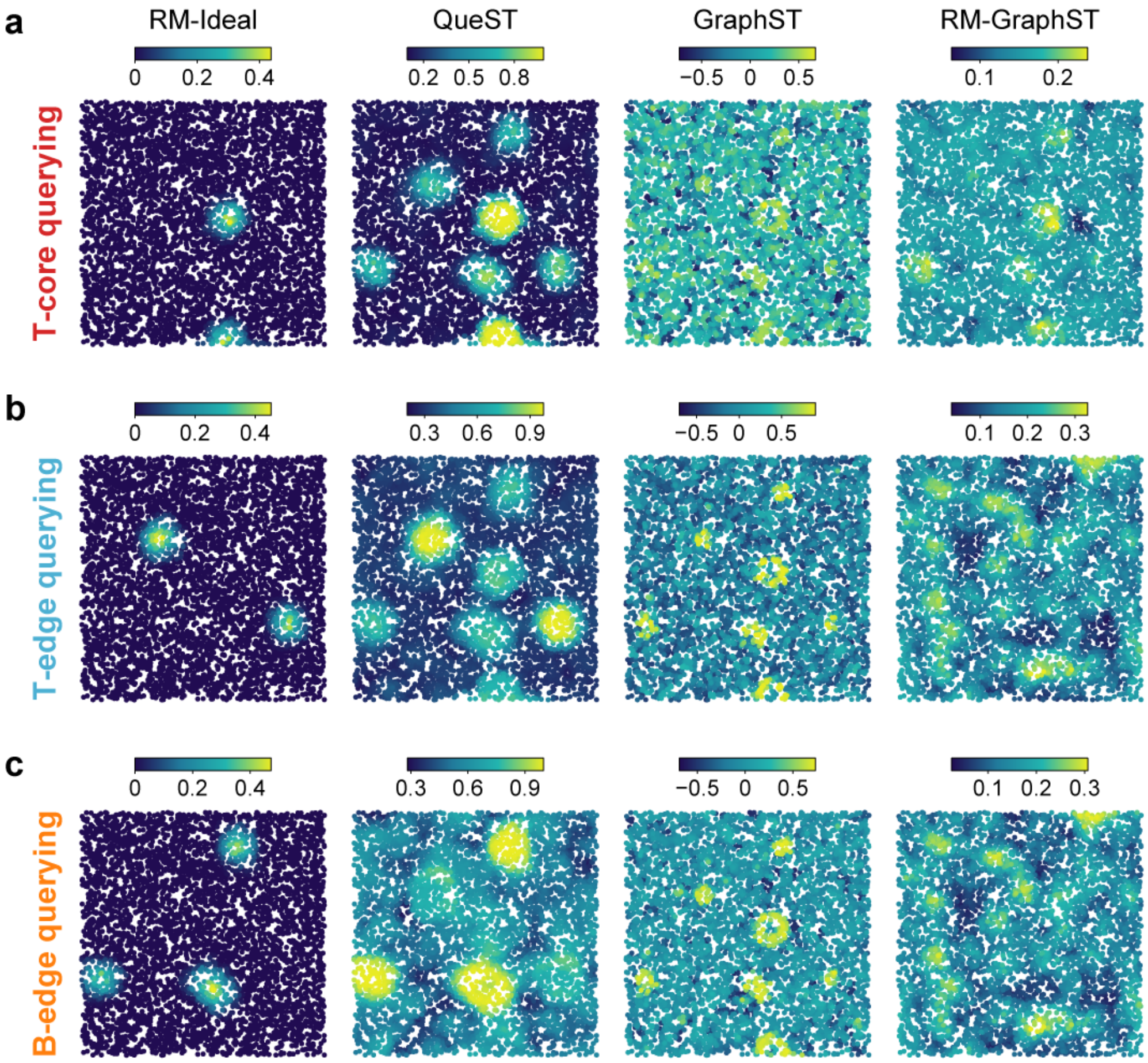

**Fig. S1.** Predicted matching scores of additional baseline methods for T-core, T-edge and B-edge querying.

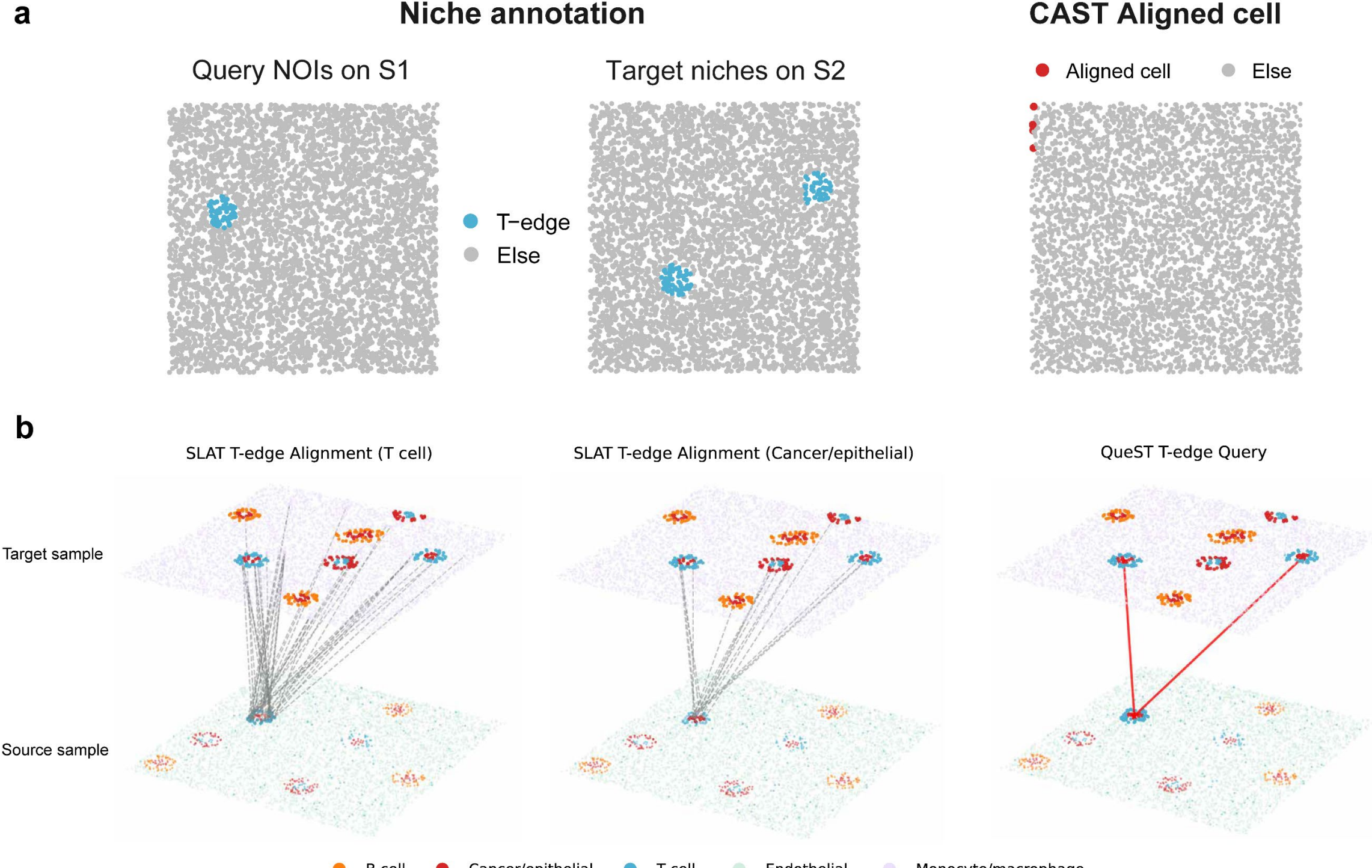

**Fig. S2. a.** Ground truth niche annotation and visualization of CAST alignment results for T-edge querying. **b.** Conceptual illustration of distinction between cell-level spatial alignment and niche-level querying using T-edge query on the simulation data as an example. Left and middle: SLAT alignment results for the T-edge query NOI. Right: QueST query results for the same T-edge NOI.

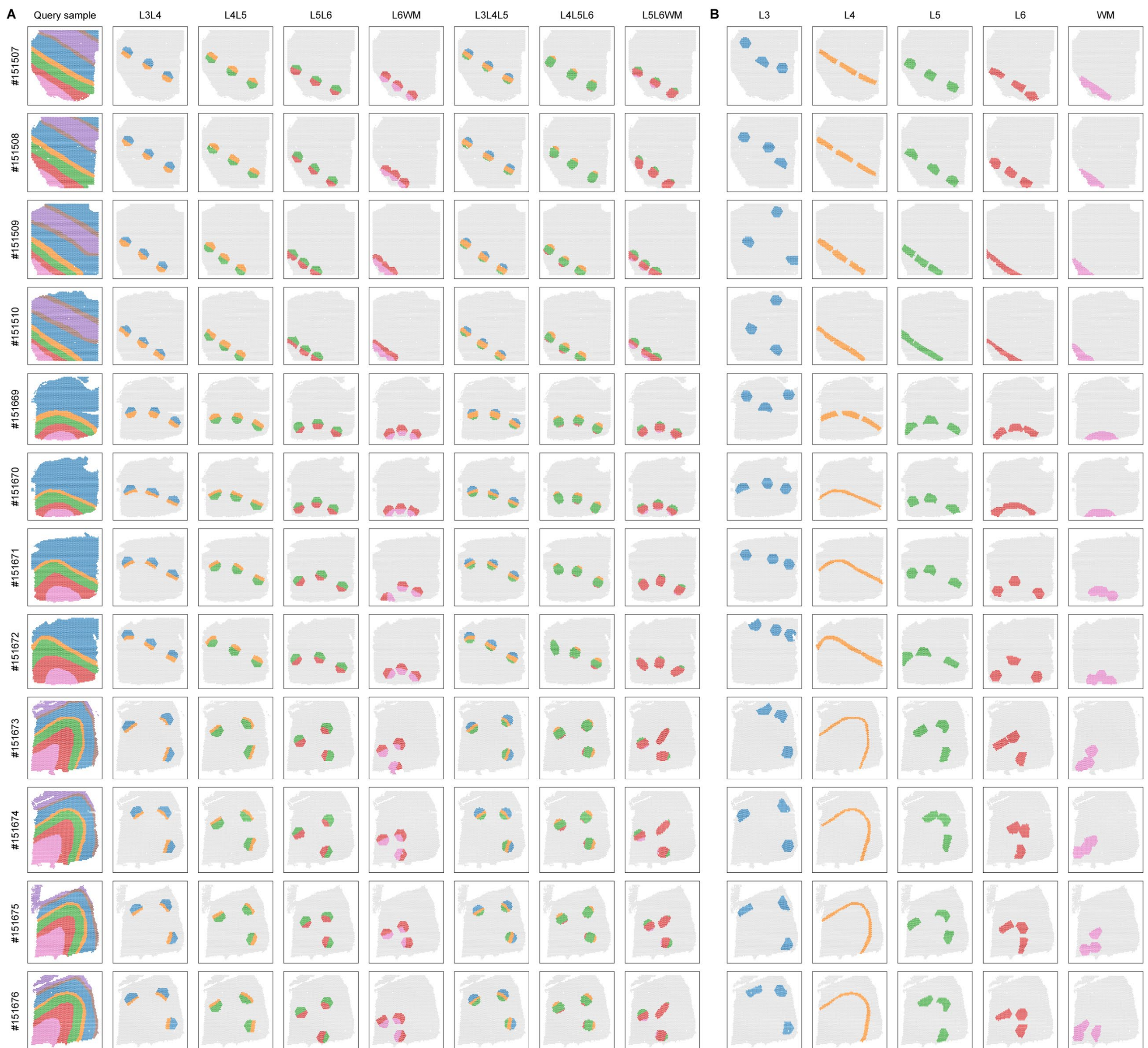

**Fig. S3. a.** Ground truth annotation and query NOIs at the boundaries between tissue layers on the DLPFC dataset. **b.** Query NOIs within homogeneous regions on the DLPFC dataset.

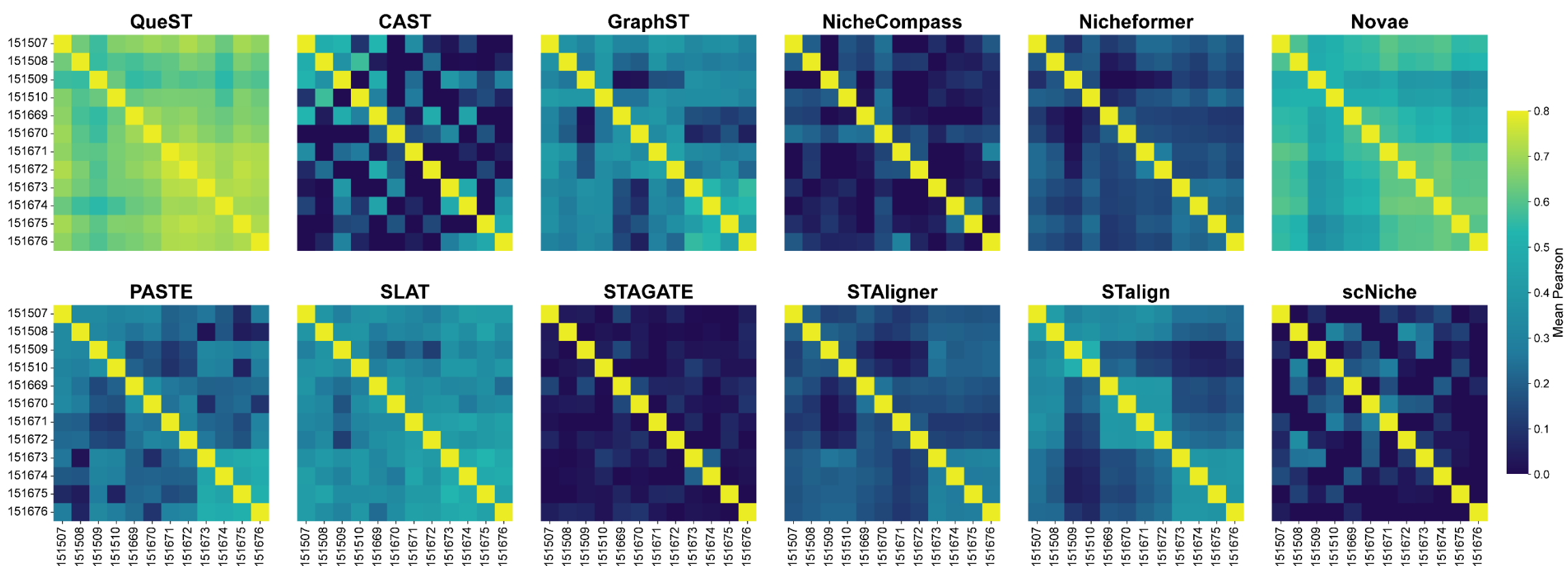

**Fig. S4.** Performance of niche querying on the DLPFC dataset in the form of sample-pair-wise PCC. Each block is colored by the mean PCC of the corresponding query-reference sample pair, with the diagonal filled with default values since we did not perform self-querying.

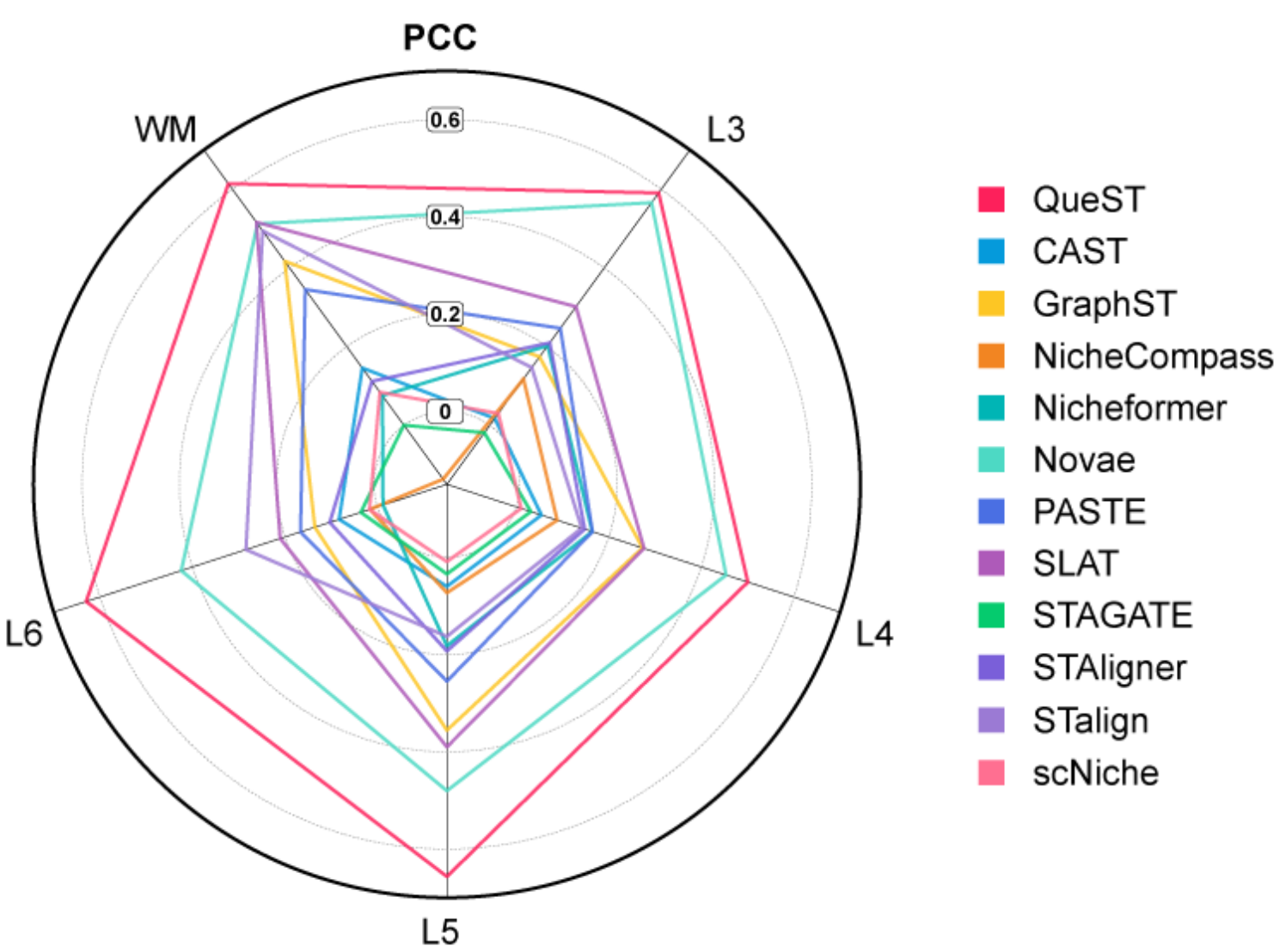

**Fig. S5.** Radar plot showing different methods' performance in the form of niche-type-wise PCCs, including both heterogeneous and homogeneous query NOIs.

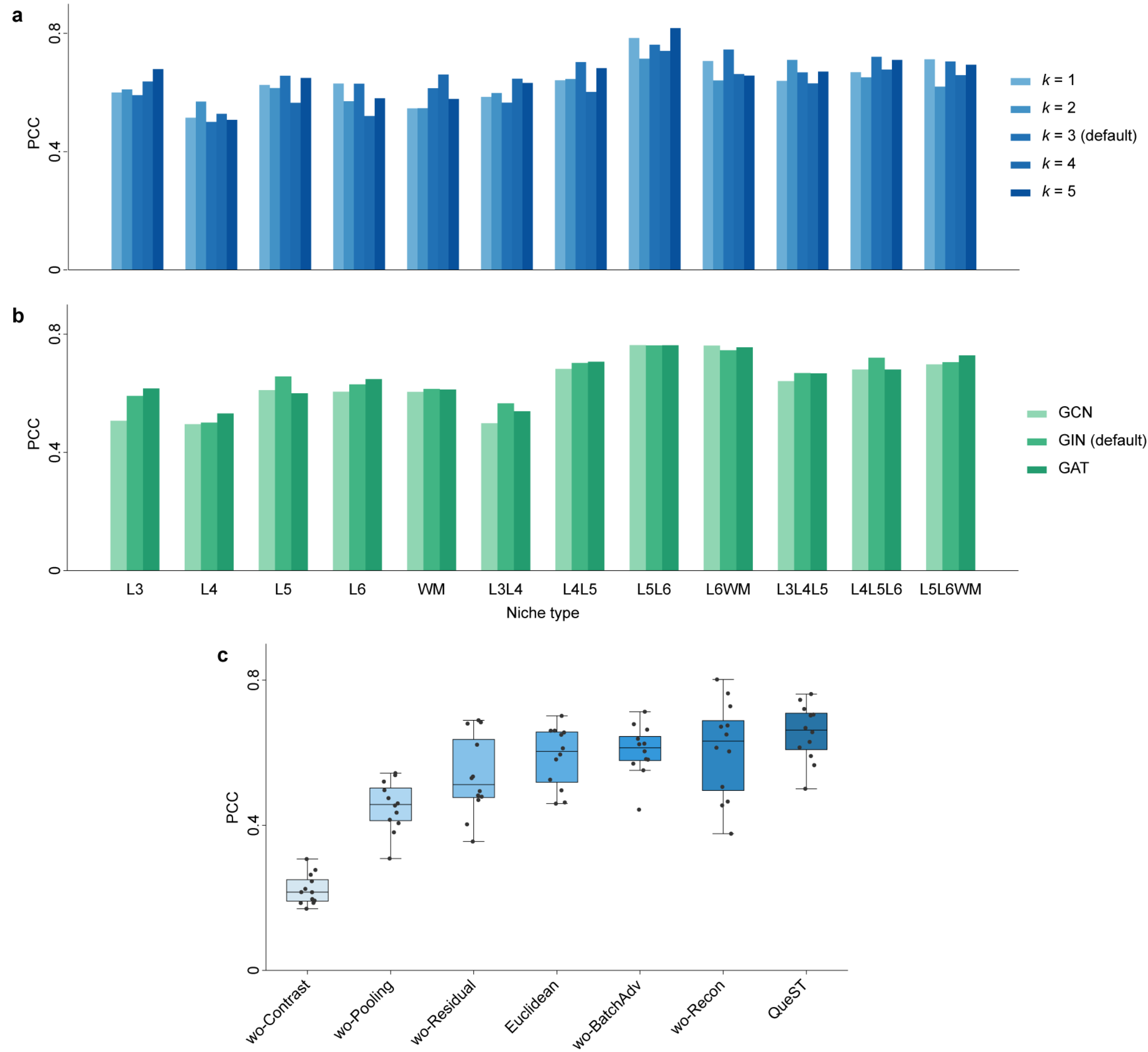

**Fig. S6. a.** Sensitivity analysis of the subgraph hop number $K$ on the DLPFC dataset. Bar plot showing the niche-type-wise PCC between QueST's predicted matching scores and RM-Ideal for $K$ = 1, 2, 3 (default), 4, and 5 across all 12 NOI types. **b.** Sensitivity analysis of the GNN backbone architecture on the DLPFC dataset. Bar plot showing the niche-type-wise PCC between QueST's predicted matching scores and RM-Ideal for GCN, GIN (default), and GAT encoders across all 12 NOI types. **c.** Ablation study of QueST's key components on the DLPFC dataset. Box plot showing the niche-type-wise PCC between each variant's predicted matching scores and RM-Ideal across all 12 NOI types, where each dot corresponds to a niche type.

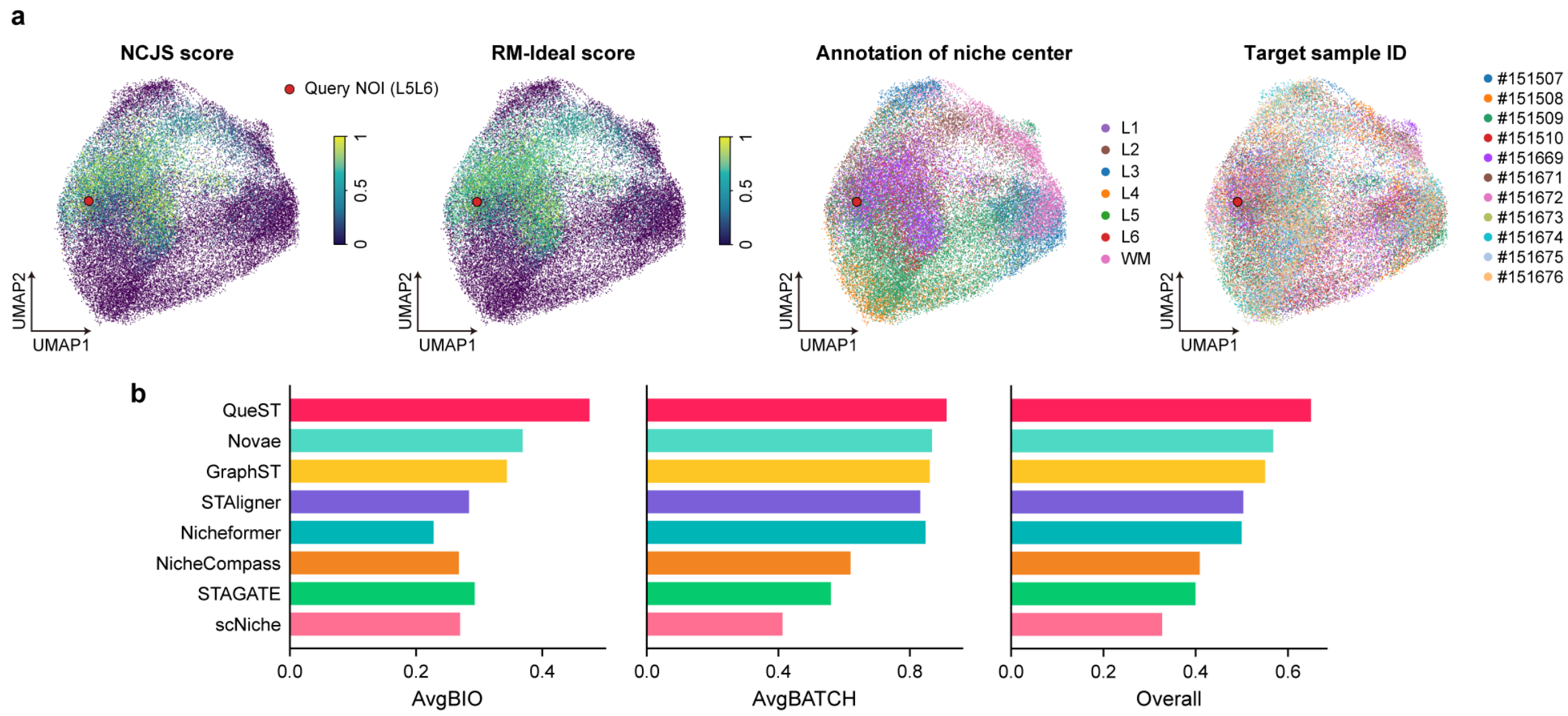

**Fig. S7. a.** UMAP visualization of GraphST's niche latent representation colored by the NCJS score, RM-Ideal score, layer type annotation of the niche center, and sample ID. **b.** Batch Integration performance on the DLPFC dataset. Performance was evaluated using the AvgBIO, AvgBATCH, and Overall score. The AvgBIO score was computed as the average of Adjusted Rand Index (ARI), Normalized Mutual Information (NMI), and Average Silhouette Width (ASW) for cell type. The AvgBATCH score was calculated as the average of ASW for batch and graph connectivity. The overall score was defined as a weighted sum: $0.6 \times$ AvgBIO + $0.4 \times$ AvgBATCH.

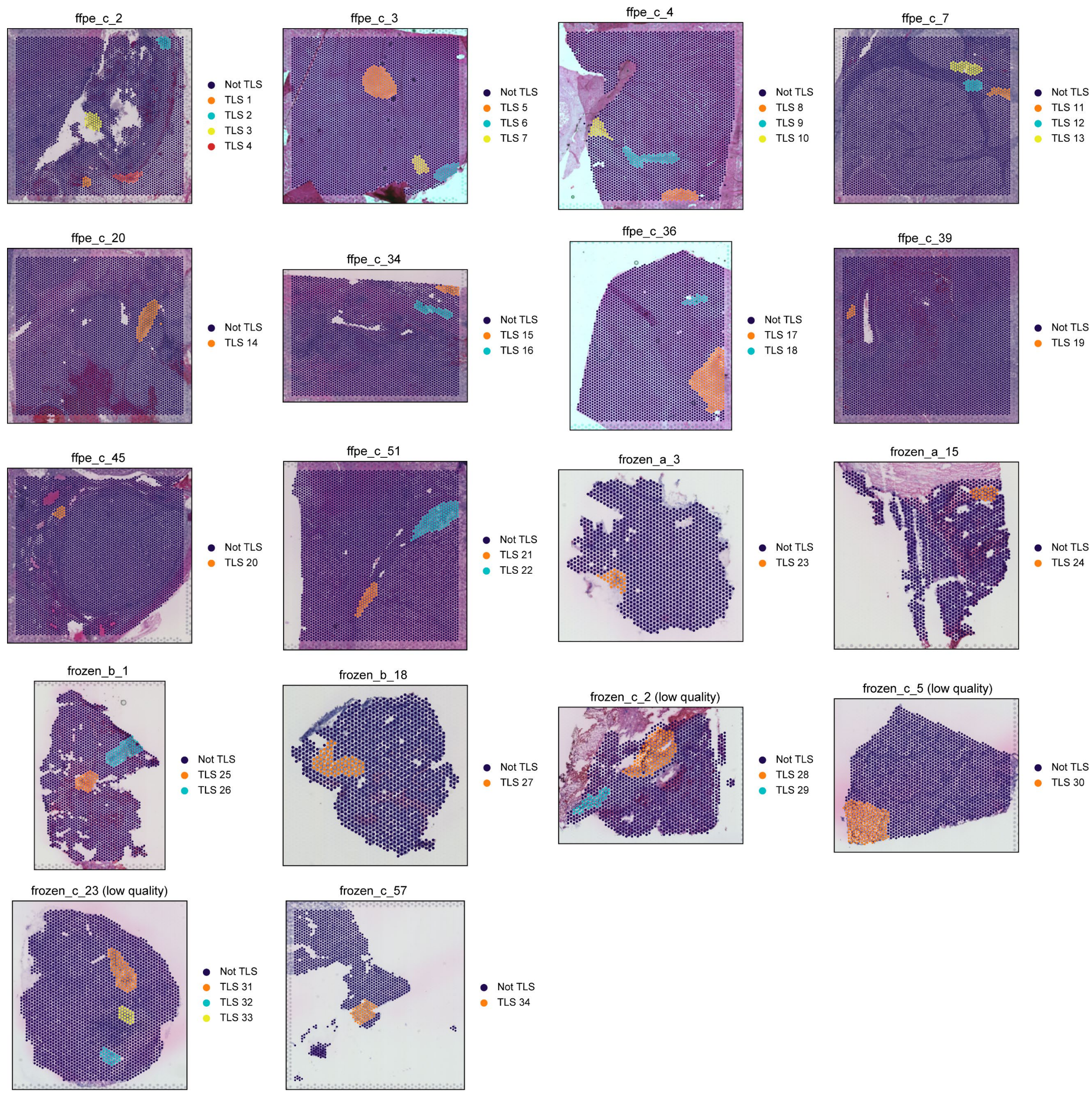

**Fig. S8.** Visualization of TLS on all the 18 TLS-positive samples of the ccRCC dataset, with frozen_c_2, frozen_c_5, and frozen_c_23 marked as low quality.

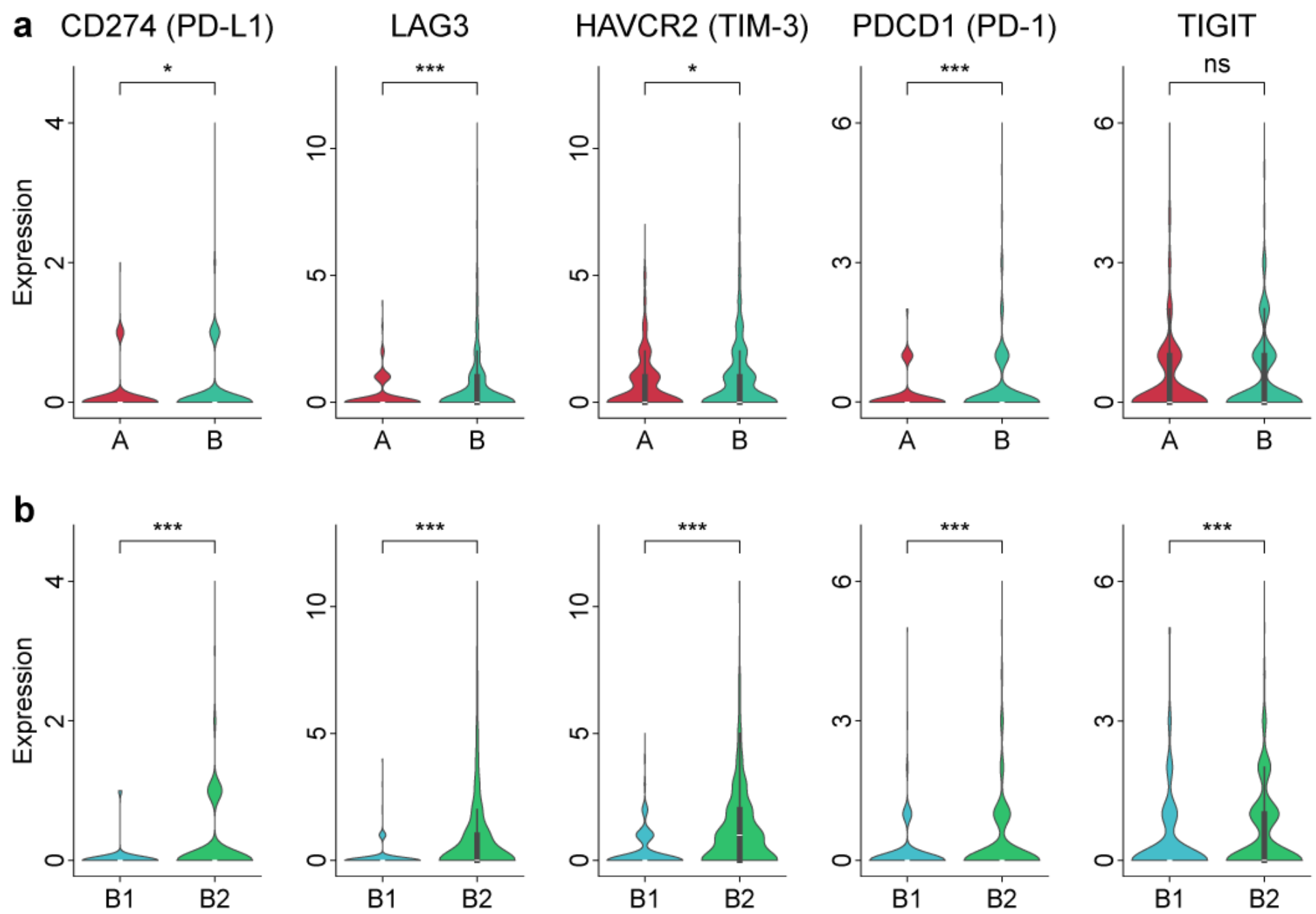

**Fig. S9. a.** Violin plots comparing the expression of immune checkpoint genes (CD274/PD-L1, LAG3, HAVCR2/TIM-3, PDCD1/PD-1, and TIGIT) between Type-A and Type-B TLSs. Asterisks denote statistical significance (* $p < 0.05$, *** $p < 0.001$). **b.** Violin plots comparing the expression of immune checkpoint genes between Type-B1 and Type-B2 TLSs.

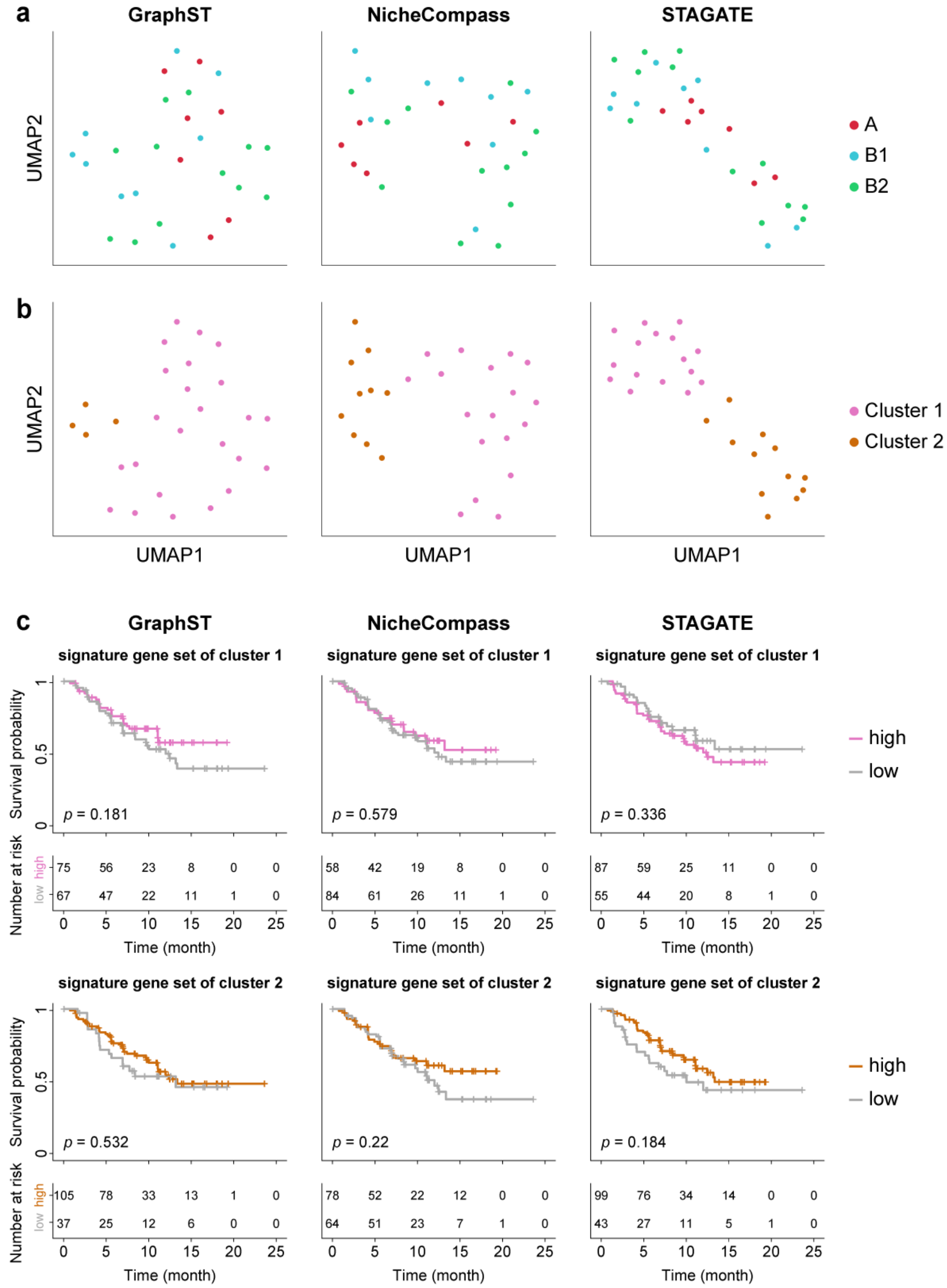

**Fig. S10. TLS subtyping of GraphST, NicheCompass, and STAGATE. a.** UMAP visualization of TLS embeddings obtained from these methods' latent space, colored by TLS subtype revealed by QueST. **b.** UMAP visualization of TLS embeddings obtained from these methods' latent space, colored by Leiden clustering labels on their embeddings. **c.** Kaplan–Meier visualizations of overall survival among the JAVELIN 101 samples stratified by ssGSEA scores based on the top 10 upregulated genes of TLS clusters obtained from Leiden clustering.

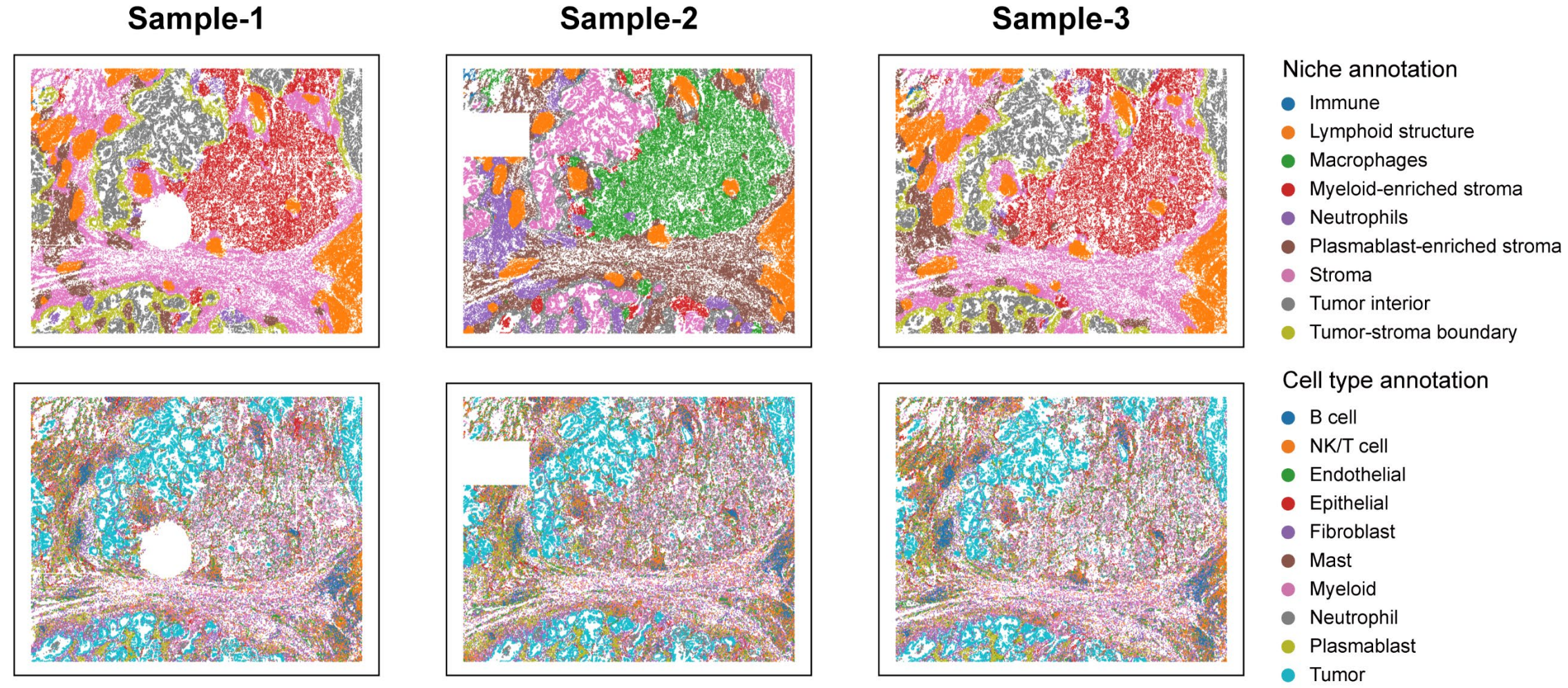

**Fig. S11.** Ground truth niche and cell type annotation of the NSCLC samples.

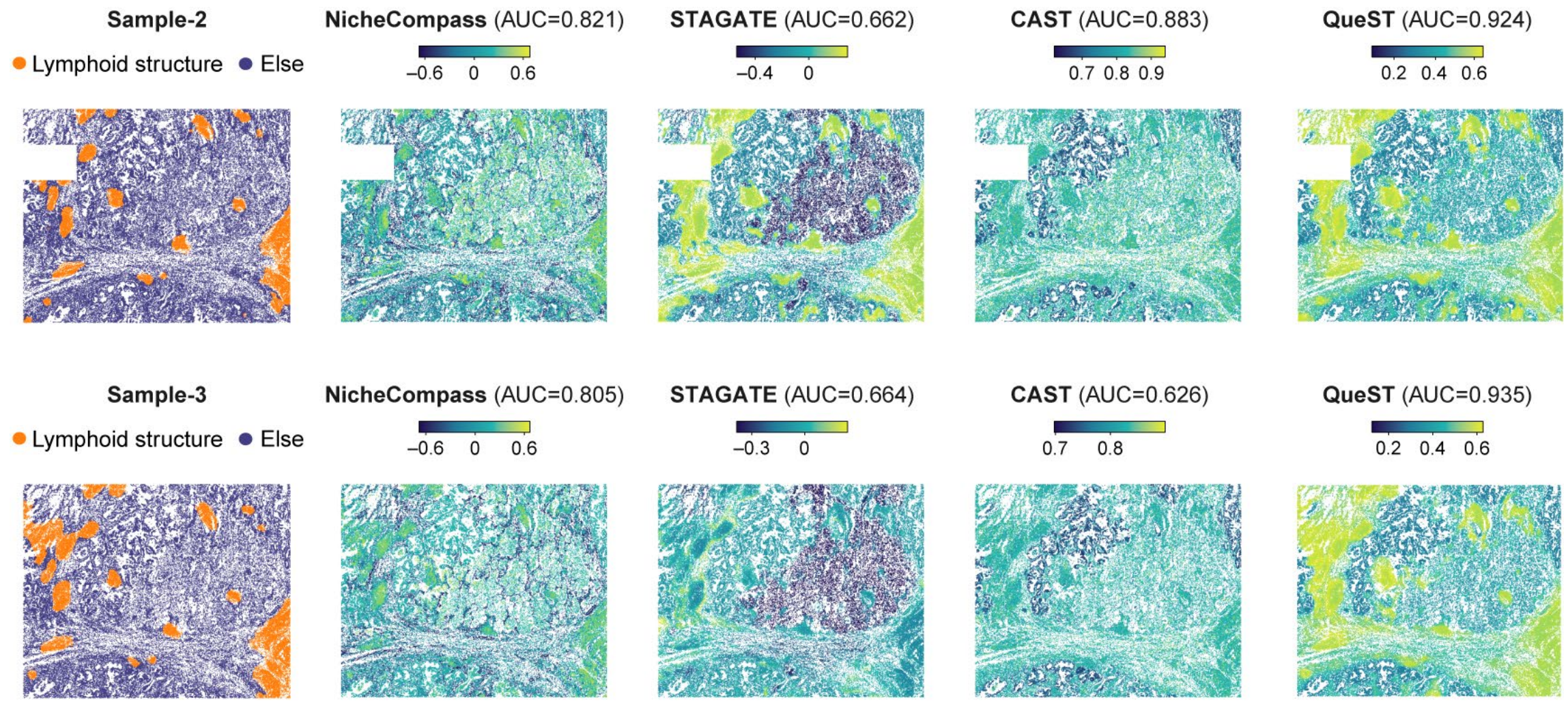

**Fig. S12.** Ground truth annotation and different methods' predicted matching scores on NSCLC Sample-2 and Sample-3.

## Supplementary Tables

**Table S1. Settings of seven query NOI types on the DLPFC dataset.**

| NOI type | Location setting | Number of spots |
|---|---|---|
| **L3L4** | Niche centered on the margin of Layer3 and Layer4 | 100 |
| **L3L4L5** | Niche with center in Layer4 but extended into Layer3 and Layer5 | 100 |
| **L4L5** | Niche centered on the margin of Layer4 and Layer5 | 100 |
| **L4L5L6** | Niche centered in Layer5 but extended into Layer4 and Layer6 | 100 |
| **L5L6** | Niche centered on the margin of Layer5 and Layer6 | 100 |
| **L5L6WM** | Niche centered in Layer6 but extended into Layer5 and WM | 100 |
| **L6WM** | Niche centered on the margin of Layer6 and WM | 100 |